\documentclass[useAMS,usenatbib,referee]{mn2e}
\onecolumn
\usepackage{graphicx}
\usepackage{natbib}
\usepackage{times}
\usepackage{float}
\usepackage{amsmath}
\usepackage{amssymb}
\usepackage{epsfig}
\usepackage{xcolor}
\definecolor{red}{rgb}{1,0,0}
\definecolor{RED}{rgb}{1,0,0}
\definecolor{blue}{rgb}{0,0,1}
\definecolor{maj}{rgb}{1,0,1}

\definecolor{white}{cmyk}{0,0,0,0}

\title[On tidal evolution of a stellar rotation axis]{On the non-dissipative tidal evolution of the misalignment between spin and orbital angular momenta }
\author[ P. B. Ivanov and J. C. B. Papaloizou ]{P.B.Ivanov$^{1}$\thanks{E-mail:
pbi20@cam.ac.uk (PBI)}, J. C. B. Papaloizou $^{2}$\thanks{E-mail:
J.C.B.Papaloizou@damtp.cam.ac.uk (JCBP)}\\
$^{1}$Astro Space Centre, P.N. Lebedev Physical Institute, 84/32
Profsoyuznaya Street, Moscow, 117997, Russia  \\
$^{2}$ DAMTP, Centre for Mathematical Sciences, University of
Cambridge, Wilberforce Road, Cambridge CB3 0WA }

\begin{document}

\date{Accepted. Received; in original form}

\pagerange{\pageref{firstpage}--\pageref{lastpage}} \pubyear{2010}

\maketitle

\label{firstpage}

\begin{abstract}
{ We extend our previous work
on the evolution of
close binary systems with misaligned orbital and spin angular momenta resulting from non-dissipative tidal interaction
to include all physical effects contributing to apsidal motion.
In addition to tidal distortion of the primary by the compact secondary these 
include  relativistic Einstein precession and the rotational distortion of the primary.
The influence of the precession of the line of nodes is included.
The dependence of the tidal torque on the  apsidal  angle $\hat\varpi$ couples 
the apsidal motion to the rate of evolution of the misalignment angle $\beta$ which is 
 found to oscillate. We provide analytical estimates for the oscillation amplitude $\Delta\beta$
 over a wide range of parameter space confirmed by numerical integrations. This
is found to be more significant near critical curves on which $d{\hat \varpi } /dt=0$ for a specified $\beta.$

We find that to obtain  $0.1 < \Delta\beta < \sim 1,$
  the mass ratio, $q  > \sim1$ the initial eccentricity should be modest,
$\cos \beta < 1/\sqrt{5},$ with $\cos\beta <0 $ corresponding to retrograde rotation,  initially,  and the primary rotation rate should be sufficiently large.
The extended discussion of apsidal motion and its coupled evolution to the misalignment 
angle given here has potential applications to close binaries with anomalous apsidal motion as
 well as  transiting exoplanets such as warm Jupiters.}

\end{abstract}


\section{Introduction}

In binary and exoplanetary system there could be a situation when rotational axis of a companion is inclined with
respect to orbital plane. Recently, this possibility has received observational confirmation, see e.g. \cite{Alb} for a discussion
of this effect in case of binary system DI Herculis and \cite{Alb1} and references there in for a discussion of exoplanetary systems 
with close-in planets on orbits inclined with respect to rotational axis of the parent star.
{In addition the two transiting warm Jupiters on eccentric orbits, TOI 5152b and TOI-5153b,  could 
potentially exhibit such a misalignment \citep{UM2022}.

For sufficiently small separation of the components of  a  binary/exoplanetary system, tidal interaction may play a significant role in governing orbital evolution  \citep[see e.g.][for a general recent discussion]{O2014, Barker2020}.
When there is a misalignment  between the rotation axis and the orbital angular momentum tidal interactions are significantly modified in comparison to the more frequently studied aligned case,
\citep[see e.g.][]{EKH,Barker2009}.
  
A seminal theory of the  quasi-static tidal interaction between gaseous objects on inclined orbits, valid for any value of the angle of  inclination between the rotation axis and the orbital angular momentum was proposed
by \cite{EKH}. Recently, \cite{IP}, hereafter IP,  revised the theory of \cite{EKH}, incorporating Coriolis forces and a self-consistent treatment of energy  dissipation based on first principles. This made use of a formalism previously applied
to dynamics tides  \citep[see e.g.][]{IP7,IPCh}. They avoided neglecting Coriolis forces  as well as making any ad hoc assumptions on the character of the tidal interaction, and the energy dissipation rate, 
as was done in \cite{EKH}.

IP found that qualitatively  new effects arise from the consideration of Coriolis and inertial forces.  
Their scale is  proportional to stellar rotation frequency $\Omega_r$. A consequence is evolution of the inclination
angle, $\beta,$ together with the orbital angular momentum, in the regime in which energy  is conserved
(the non dissipative regime).
In this regime both orbital and rotational energies and, accordingly, the
orbital semi-major axis $a,$ and $\Omega_r$,are conserved.

As discussed in IP the physical origin of such non-dissipative evolution is
associated with the Coriolis and inertial forces generating a tidal response displacement that has an angular dependence differing from that of the tidal forcing which it would otherwise take.
The effect  can be regarded as acting in a  similar way  to  the well-known Lidov-Kozai effect, but, in our case there is no need for the presence of a third body to cause the joint evolution of the orbital eccentricity and the orbital angular momentum. 

It is important to note that due to the inefficiency of dissipative 
processes operating in gaseous celestial bodies the corresponding characteristic   time scales  of evolution are typically 
very long compared to those associated with non dissipative evolution.
Moreover, e.g. turbulent viscosity, which may lead to dissipation of quasi-static tides in many potentially interesting objects may be too weak to be important, see e.g. \cite{Dug} and references therein.

For non-dissipative evolution the corresponding torque acting between the primary and orbit  is proportional to $\sin 2{\hat\varpi}$, where ${\hat\varpi}$ is the angle characterising 
the orientation of the  apsidal line of the orbit. Therefore, the characteristic time scale of  evolution is in part determined by the rate of apsidal precession.
This may  have several different sources. 
IP considered the  situation where tides are  exerted only on component (the primary star), the secondary being compact. 
Also, they took into account  only classical  apsidal precession induced by tidal distortion. 
They found that the inclination angle exhibited periodic motions with  period one half of the period of apsidal precession. 
The amplitude  was  determined by several factors, most importantly,  $\Omega_r$, $a$, the orbital eccentricity $e,$ the  stellar moment of inertia $I,$ and the mass ratio $q$ between the secondary and primary. 

In this paper we generalise results of IP considering taking account of all expected contributions to apsidal motion for a binary of the type we consider. These include  relativistic Einstein precession and effects arising from the flattening of the primary due to its rotation, see e.g. \cite{BOC}. It is important to
note that the latter effect depends on the inclination angle $\beta$ and may change sign.  This dependence was used by \cite{Sh} 
to explain an unusual apsidal motion of DI Herculis and, later, was invoked to explain properties of AS Camelopardalis, see \cite{Pavl}.
It is also important to note that the orientation of the  apsidal angle is made with respect to the line of nodes which is also precessing, a feature that
also depends on $\beta.$ This will affect the rate of precession of the apsidal line that we require. 
Thus, when all these effects may  play a significant role,  a coupled evolution of the angles  $\beta$ and ${\hat\varpi}$ is  expected.

We analyse in detail qualitative properties of the resulting dynamical system, which describes the evolution of  $\beta$ and 
${\hat \varpi},$ with the  orbital eccentricity $e$ being determined as a dependent quantity.
 We begin by  providing conditions, under which any one process gives the dominant contribution to apsidal motion, going on
to  estimate a typical magnitude for the  expected change  to the  inclination, $\Delta \beta,$ in each case.

We go on to consider  the situation when the system, in the course of its evolution crosses a 'critical curve', 
in the parameter space of the problem, defined by the condition, that the total apsidal precession rate is zero
for a particular value of $\beta$, namely $\beta_0$.  The discussion given here is expected  be useful
for assessing the possibility of dramatic reductions or reversals in the direction of apsidal motion in close binary systems similar to DI Herculis.

In this situation it is expected that amplitude of  variation, $\Delta \beta$ is much larger than for the previous case.
We discuss in detail  the properties of such 'critical curves' finding  that one can only be crossed  when $\beta_0$ is  relatively large and  possibly corresponding to  retrograde rotation such that $\cos \beta_0 < {1/ \sqrt{5}}$.

We study the evolution of $\Delta \beta$ when the system evolves near a critical curve making  the assumption that the magnitude of  $\Delta \beta $ is small. We show that it is formally governed by a simple  pendulum equation.
Is found that the system's behaviour changes drastically for such solutions. The apsidal angle changes periodically 
(librates), while variations of $\Delta \beta $ can be large enough to lead to periodic changes in $\beta$ corresponding to switching  from prograde to retrograde rotation and back.

We confirm our analysis  by considering two numerical integrations and discuss the four conditions we found to be required in order to obtain  $\Delta \beta$  in the range $0.1-1$. These were:
1) $\Omega_r$ should be large enough, 2) the eccentricity should be moderately large, say, $e\sim 0.5$, 3) the initial inclination, $\beta_0,$ should be large enough, 4) the mass ratio $q$ should be order of unity or larger. The case of large mass ratio could, for example, be applicable to  a tidally active planet with its rotational axis strongly inclined with respect to the orbital plane. In an accompanying paper \citep{IP1} 
a larger preliminary numerical survey of parameter space also provides some further  confirmation of these conditions.}

{The effects discussed in this paper  could have several possible observational implications. The discussion of the processes contributing
to apsidal motion incorporating the precession of the line of nodes as well as of the critical curves could be applicable to future studies of transiting exoplanets in orbits with significant 
eccentricity and misalignment \citep{UM2022}.  As already noted these effects  may also be relevant to the  light curves of eclipsing binaries such as DI Herculis \citep{Sh}. In addition, significant changes in $\beta$ may be possible in such systems.
The effects studied here  may  also play a role when the system's evolution on longer dissipative  time scales is considered.

The structure of this paper is as follows. In Section \ref{Basiceq} we introduce our basic notations and definitions. In Section \ref{inclination} we discuss the basic equations governing the non-dissipative evolution of our system. In Section \ref{eveq} we provide a qualitative analysis of it and estimate the variation of $\Delta \beta$ under the assumption that a single process dominates the  apsidal precession rate.  In Section \ref{S5} we discuss the determination and  properties of the 
'critical curves'  and in Section \ref{evolution} we discuss solutions evolving close to  a critical curve  both analytically and numerically. Finally, in Section \ref{concl} we conclude by discussing the possible implications and extensions of this work.}


\section{Basic definitions and notation}\label{Basiceq}

We consider  a binary  that consists of a primary star of mass $M_{*}$  and  radius  $R_*$  together with a point-like secondary star of mass
$M_1$. The orbit of the binary is assumed to be, in general, elliptic, with eccentricity $e$ and semi-major axis $a$. 
There are three dynamical  frequencies that are significant for our purposes, a typical inverse dynamical time scale associated with the primary
 $\Omega_{*}=\sqrt{{GM_*/ R_*^3}}$,
where $G$ is gravitational constant, the mean motion $n_0=\sqrt{{G(M_*+M_1)/ a^3}}$, and the  rotation frequency of the primary star
$\Omega_r$. Is is also convenient to use the dimensionless semi-major axis $\tilde a=a/R_*$, the ratio of the rotation frequency to the orbital mean motion
$\sigma~=~\Omega_r/n_0,$ \footnote{This deviates slightly from the notation of IP in which  $\sigma=\Omega_r/(\lambda n_0)$ with $\lambda$ being defined there.
This quantity is not used in this paper.}  and the mass ratio $q=M_1/M_*$. 

The orbital angular and stellar spin angular momentum vectors are ${\bf L}$ and ${\bf S}$, respectively, their sum ${\bf J}={\bf L}+{\bf S}$
defines the total angular momentum of the system, which is conserved in the course of orbital evolution. We define inclination angles 
$\beta$, $i$ and $\delta$ as inclination angles between ${\bf S}$ and ${\bf L}$, ${\bf L}$ and ${\bf J}$ and ${\bf S}$ and
${\bf J}$, respectively, with their relative orientations chosen in such a way, that $\delta = \beta - i$ (see also IP). 

We have the obvious relations following from the definition of these angles (see IP) 
\begin{align}
&\cos {\beta}={({\bf L}\cdot {\bf S})\over LS}, 
\label{e0}
\end{align}
where $L$ and $S$ are the magnitudes  of ${\bf L}$ and ${\bf S}$, and 
we have $S=I\Omega_r$, where $I$ is primary's moment of inertia, and 
$L={qM_{*}/ (1+q)}n_0a^{2}\sqrt{1-e^2}.$ Furthermore  
\begin{align}
 &\cos {i}={({\bf J}\cdot {\bf L})\over JL}, \hspace{3mm}{\rm and}\hspace{3mm} \cos {\delta}={({\bf J}\cdot {\bf S})\over JS},
 \end{align} 
where $J$ is the magnitude  of ${\bf J}$.
In addition, we also have
\hspace{2mm}  $2{\bf J}\cdot {\bf L} =J^2+L^2-S^2 $ \hspace{2mm}  and \hspace{2mm}  $2{\bf L}\cdot {\bf S} =J^2- L^2-S^2$ 
and, accordingly, the cosines of  $\beta$ and $i$ are given by,
\begin{equation}
\cos \beta ={J^2-L^2-S^2\over 2LS}\hspace{3mm}{\rm and} \hspace{3mm}\cos i = {J^2+L^2-S^2\over 2JL}.
\label{e1}
\end{equation}
From the first of these we obtain
{\begin{align}
\frac{J}{L}=\frac{1}{\sqrt{1-S^2\sin^2\beta/J^2}-S\cos\beta/J}.\label{JoverL}
\end{align}}
We can also  express the sines of $\beta$, $i$ and $\delta$ in terms of $J$, $L$ and $S$, thus obtaining
\begin{align}
&\sin \beta =\frac{\sqrt{(J^2-(L-S)^2)((L+S)^2-J^2)}}{2LS},\hspace{2mm} {\rm and}\nonumber\\
&\hspace{0mm} \sin i = \frac{\sqrt{(S^2-(J-L)^2)((J+L)^2-S^2)}}{2JL}.           
\label{e2}
\end{align}
In addition, consideration of the angular momentum components perpendicular to
{\bf J} and {\bf S} respectively gives
\begin{equation}
\sin \delta ={L\over S}\sin i = {L\over J} \sin \beta.
\label{e3}
\end{equation}

It is clear that vectors ${\bf L}$, ${\bf S}$ and ${\bf J}$ lie in the same plane. For our purposes, it is useful to introduce two orthonormal 
right oriented triads of unit vectors, defining two Cartesian coordinate systems $(X,Y,Z)$ and $(X^{'},Y^{'},Z^{'})$ in such a way, that {the $Y$ and $Y'$ axes are colinear} and  lie
in the direction perpendicular to this plane, {while the $Z$ and  $Z'$ axes are directed  along 
${\bf S}$ and ${\bf L}$, respectively.} From these definitions and {the above discussion} it follows that {we can choose} the first triad ${\bf e}_x, {\bf e}_y, {\bf e}_z$
to be explicitly represented in the form
\begin{equation}
{\bf e}_{x}={({\bf s}\times {\bf j})\times {\bf s}\over \sin \delta}= {{\bf j}-\cos \delta {\bf s}\over
\sin \delta}, \quad {\bf e}_{y}={{\bf s}\times {\bf j}\over \sin \delta},
\quad {\bf e}_{z}={\bf s},    
\label{e4}
\end{equation}
where ${\bf s}={\bf S}/S$ and  ${\bf j}={\bf J}/J$, while the second one  ${\bf e}_{x^{'}}, {\bf e}_{y^{'}}, {\bf e}_{z^{'}}$ can be obtained 
from (\ref{e4}) by the substitution $\delta \rightarrow  i$ and ${\bf s} \rightarrow {\bf l}$, where ${\bf l}={\bf L}/L$:    
\begin{equation}
{\bf e}_{x^{'}}={({\bf l}\times {\bf j})\times {\bf l}\over \sin i}={{\bf j}-\cos i {\bf l}\over
\sin i} , \quad {\bf e}_{y^{'}}={{\bf l}\times {\bf j}\over \sin i}, 
\quad {\bf e}_{z{'}}={\bf l}.    
\label{e5}
\end{equation}  
Later on we are going  to call coordinate frames defined with help of (\ref{e4}) and (\ref{e5}) as 'stellar' and 'orbital' frames, respectively

\section{Equations governing the non-dissipative tidal evolution of the  inclination
angle between the spin and orbital angular momentum vectors}
\label{inclination}

In order to discuss the non-dissipative evolution of the inclination angles we need  to relate non-dissipative contribution to the tidal torque acting in the stellar frame, which was  provided in IP,  to time derivatives of these angles. 
This can be easily done by differentiating $\cos \delta = ({\bf j}\cdot {\bf s})$ with respect to time, taking into account that ${\bf j}$ is conserved and 
with the  help of eq. (\ref{e4}) 
expressing
${\bf j}$ in terms of ${\bf e}_{x}$ in the resulting expression, thus we obtain 
\begin{align}
\dot \delta =-{\sin \delta ({\bf e}_x\cdot \dot {\bf s})+\cos \delta
 ({\bf s}\cdot \dot {\bf s})\over \sin \delta}. 
 \end{align}
But, $({\bf s}\cdot \dot {\bf s})=0$, so we have
 
\begin{align}
\dot \delta =-{T^{x}\over S},
\label{e6}
\end{align}
where $T^x= S({\bf e}_x\cdot \dot {\bf s})=({\bf e}_x\cdot {\dot {\bf S}})
\equiv({\bf T}\cdot  {\bf e}_x)$ is the component of the torque ${\bf T}$  in the $X$ direction acting on the star.

Derivation of the evolution equation for the angle $\beta $ proceeds in a similar way. We first  differentiate  equation (\ref{e0}) with respect to time. We then note    
that angular momentum conservation implies that $\dot {\bf L}=-\dot {\bf S}$ and
we use the fact that for non dissipative 
evolution, $( {\bf T}\cdot  {\bf e}_z) =0, $ so that, $S$ is conserved (IP),
and, accordingly,
 $({\bf S}\cdot \dot {\bf L})=~-({\bf S}\cdot \dot {\bf S})~=~0$. In this way we obtain
\begin{align}
\dot \beta =-{1\over \sin \beta}{({\bf L}\cdot \dot {\bf S})\over L^2}\left({L\over S}+\cos \beta\right).
\label{e7}
\end{align}  
From eq.(\ref{e4}) we find in addition that, $\sin\delta ({\bf e}_x\cdot{\dot{\bf S}})= ({\bf L}\cdot \dot{\bf S})/J.$ 
We  then use (\ref{e3}), thus obtaining
\begin{align}
\dot \beta=-\left( \frac{1}{S}+ \frac{\cos\beta}{L} \right) T^x 
\label{e8}
\end {align}
( see also equation (17) of IP). An evolution equation for the angle $i=\beta - \delta$ can be easily obtained from (\ref{e6}) and 
(\ref{e8}). 

\noindent In addition we have  the conservation of  the total angular momentum which yields
\begin{align}
L^2/(2S)+L\cos\beta = (J^2-  S^2)/(2S) = {\rm constant},\label{integral}
\end{align} 
where we recall  that $S$ is constant.

\subsection{An explicit expression for $T^x$}
The torque component $T^x$ requires an extensive analysis
which is carried out in IP. The reader is referred there for details.
Here we note that in the equilibrium tide approximation for a barotropic stellar model
of the type we consider, $T^x=0$.
However, a non zero value is obtained when the induced acceleration and effective
Coriolis force is included in the determination of the tidal response.
This  is carried out in Section 5 of  IP {with some discussion of the origin of a non zero value of $T^x$ given in Section 5.4.2.}
 The results are then used to obtain $T^x$ 
in Section 6 and Appendix B of IP.
{\subsection{Quasi-static and dynamical tides}\label{eqdyntide}
IP considered the density response,  $\rho_{n,k},$ to the perturbing potential
${\overline U}=  r^2  {\cal A}_{n,k}Y_{2,n}(\theta, \phi),$
where the forcing frequency is $\omega_f = kn_o+n\Omega_r,$
with $n$ being the azimuthal mode number and $k$ an integer.
For the definition of other quantities here and in the rest of this Section see IP.
The associated displacement is 
$\mbox{{\boldmath$\eta$}}=\mbox{{\boldmath$\xi$}}_{eq,n,k}+\mbox{{\boldmath$\xi$}}_{eq1,n,k},$
This is written as the sum of two parts, $\mbox{{\boldmath$\xi$}}_{eq,n,k},$ identified as the equilibrium, or  quasi-static,  tidal displacement
and $\mbox{{\boldmath$\xi$}}_{eq1,n,k}$ which is the difference between the displacement and that quantity.
The latter incorporates the dynamical tide.
The quantity, $\mbox{{\boldmath$\xi$}}_{eq,n,k},$ can be  taken to be the displacement in the limit of zero forcing frequency. This
together with   $\mbox{{\boldmath$\eta$}}$
 was specified by equations (57) and (27) of IP to be  in  a  purely spheroidal form.
However, it is important to note that the analysis given Sections 5.2 and 5.3 of IP  does not
depend on this assumption. 

Furthermore the analysis, aimed at specifying the overlap integral,
$\int\rho_{n,k}^{'*}  r^2Y_{2,n}(\theta, \phi)dV,$ this being required in order to determine the tidal torque,
 can be undertaken while retaining
$\mbox{{\boldmath$\xi$}}_{eq1,n,k}.$  One finds that equation (53) of IP specifying the overlap integral is retained 
but with modified definitions of the quantities, $\beta_*$ and $\Gamma$ defined in IP, the latter being neglected for the
non dissipative evolution considered here.  Hence,  \hspace{2mm}  $ n\omega_f^2(\beta_*-1)\Omega_r \rightarrow$
\begin{align}
&-\omega_f^2{\cal R} \left(
\frac{\int {\rm i} \rho\Omega_r\mbox{{\boldmath$\xi$}}_{eq,n,k}^*\cdot( {\bf \hat{k}}\times\mbox{{\boldmath$\eta$}}) dV}
{\int\rho|\mbox{{\boldmath$\xi$}}_{eq,n,k}|^2dV} -\frac{\omega_f}{2}\left(
\frac{\int  \rho\mbox{{\boldmath$\xi$}}_{eq,n,k}^*\cdot\mbox{{\boldmath$\xi$}}_{eq1,n,k}dV}
{\int\rho|\mbox{{\boldmath$\xi$}}_{eq,n,k}|^2dV}- \frac{\int  \rho\mbox{{\boldmath$\xi$}}_{eq,n,k}^*\cdot\mbox{{\boldmath$D$}}_{NA} dV}
{\omega_f^2\int\rho|\mbox{{\boldmath$\xi$}}_{eq,n,k}|^2dV}\right)
\right),
\label{EDs4}
\end{align}
where ${\cal R}$ indicates the real part, ${\bf D}_{NA}$ represents the dissipative terms in equation (35) of IP,
and for completeness $\Gamma\rightarrow \Gamma {\int\rho|\mbox{{\boldmath$\eta$}}|^2 dV}/({\int\rho|\mbox{{\boldmath$\xi$}}_{eq,n,k}|^2 dV}).$
IP then assume that $\mbox{{\boldmath$\xi$}}_{eq1,n,k}$ can be neglected in comparison to $\mbox{{\boldmath$\xi$}}_{eq,n,k}$ so that
in addition  $\mbox{{\boldmath$\eta$}} \rightarrow \mbox{{\boldmath$\xi$}}_{eq,n,k}$ in (\ref{EDs4}), which becomes
the same as equation (55) of IP, when non adiabatic  effects which are assumed to be weak in comparison to conservative effects are neglected.

IP discuss the evaluation of $\beta_*$ in this case using the form of the equilibrium tide given by equation (57) of IP.
This spheroidal form applies in the case of a non rotating spherical star and is such that $\beta_*$ and $\omega_{eq}$
are constants independent of $n.$ However, an alternative form for the equilibrium tide could be adopted with corresponding change
to  $\mbox{{\boldmath$\xi$}}_{eq1,n,k}.$ Then IP, as well as the discussion below, effectively make the approximation of adopting  constant values for
 $\beta_*$ and $\omega_{eq}$ independent of $n.$ Note that Coriolis forces are not necessary to obtain a non zero value of $\beta_*.$
 If they are neglected $\beta_*=1.$ 
 
 However, neglecting  $\mbox{{\boldmath$\xi$}}_{eq1,n,k}$ and adopting equation (57) of IP
 neglects the possibility of resonances due to eg. inertial modes \citep[see eg.][]{Papaloizou2005, O2014} or  $r$ modes \citep[see eg. ][]{PapSav} which are associated with the dynamical tide.
 But the latter resonances are highly localised in parameter space and accordingly unlikely to play a significant role. 
Note too that as only the density perturbation and associated overlap integral is required to obtain tidal torque, further identification of the
form of the displacement is not needed.

\subsection{Equation governing the evolution of $\beta$}}

 Equations (90) and (92) of IP then
specify $T^x$ through
\begin{align}
&T^x= -T_*   \frac{3(2\beta_*+1)}{2} e^2(1-e^2)^{3/2}\left(1+e^2/6\right)
\left( \frac{\Omega_r}{\omega_{eq}} \right)^2       \sin\beta\sin{2{\hat \varpi}},  \nonumber \\
&\hspace{-3mm}{\rm where}\hspace{2mm}T_* = \frac{3k_2q^2}{( 1 + q)} \left(\frac {R_*}{a}\right)^5{M_{*} n_{o}^{2} a^2}{(1 - e^2)^{-6}}\hspace{3mm}{\rm and}\label{TX}
\end{align}

\noindent $\beta_*$ is a constant of order unity ( see equation (55) of IP) , $\omega_{eq}$ differs from $\Omega_*$ by numerical factor
order of unity, $k_2$ is the apsidal motion constant and {${\hat \varpi}-{\rm \pi}/2$ is the angle between the apsidal line and the $Y$ axis which may be used to define the line of nodes.
Then the angle between the apsidal line and the $X^{'}$ axis is $\varpi = {\hat\varpi} -{\rm \pi}$.}
Thus we have
\begin{align}
\dot \beta=&\left( \frac{1}{S}+ \frac{\cos\beta}{L} \right)
 T_*   \frac{3(2\beta_*+1) e^2(1-e^2)^{3/2}}{2}\times \nonumber\\
& \left(1+\frac{e^2}{6}\right)
\left( \frac{\Omega_r}{\omega_{eq}} \right)^2   \sin\beta\sin{2{\hat \varpi}}. \label{betaeq}
\end {align}

Provided that a dependence of ${\hat \varpi}$ on time is specified equations (\ref{e8}), (\ref{integral}) and (\ref{betaeq}) together with the standard 
expression of $L$ in terms of $a$ and $e$ form a complete set. We considered in IP the simplest case when apsidal precession determined 
by equilibrium tides is given by the classical expression
\begin{align}
\frac{d{\hat \varpi}}{dt}=\frac{d{\varpi}}{dt} = \frac{d{\varpi}_T}{dt} = \frac{15 k_2n_0M_1R_*^5}{M_*(a(1-e^2))^5}\left(1+{3e^2\over 2} 
+{e^4\over 8}\right),\label{Apse}
\end{align} 
\citep{St1939}.
In this paper we would like to consider a more complicated situation taking into account other potentially important sources of apsidal precession,
namely, the Einstein precession and apsidal precession determined by
 rotational flattening of the primary \citep[e.g.][]{BOC,Sh} In the 
latter case the apsidal precession rate depends on inclination of the stellar axis to the orbit, $\beta$, which results in a much richer dynamics. We
derive an expression for the apsidal precession rate due to rotational flattening in a form appropriate for our purposes from the results of
\citet{BOC} in Appendix \ref{app}, see equation (\ref{a15}). As seen from this expression there are  two contributions,
 which have physically 
different origin. The former term is directly determined by gravitational perturbation of the Keplerian point-mass potential
arising from the rotational distortion of the primary, causing apsidal precession.
The nature of the second term proportional to $\cos i,$  is 'indirect' in the following sense.
When the rotation axis of the star is inclined with respect to the orbit, interaction 
of the  tidal potential with the misaligned  axisymmetric density distribution of the rotationally flattened star leads to precession of this  axis. This, 
in turn  causes the   orbital angular momentum vector to similarly  precess in order to conserve total angular momentum. This makes the orbital frame non-inertial  inducing  corresponding Coriolis forces, which give rise to the additional apsidal precession of the orbit.

Accordingly, adding all the contributions together we have
\begin{align}
\frac{d{{\hat \varpi}}}{dt}={d \varpi_{T} \over dt}+{d\varpi_{E} \over dt}+{d\varpi_{R} \over dt}+{d\varpi_{NI} \over dt},
\label{prec}
\end{align}     
where ${d \varpi_{T} / dt}$ is given by (\ref{Apse}), 
\begin{align}
{d\varpi_{E} \over dt}={3GM_*(1+q)\over c^2a(1-e^2)}n_{0}, \label{Ein}
\end{align}
is the standard expression for the Einstein relativistic apsidal precession, $c$ is speed of light, and ${d\varpi_{R} / dt}$ and ${d\varpi_{NI} / dt}$ are given by the first and second contributions to the apsidal advance rate  specified by eq.  (\ref{a15}).

\subsection{Evolution equations in dimensionless form} \label{eqdim}

In order to simplify the discussion of the evolution equations 
we obtain a dimensionless form of equation (\ref{betaeq})) by 
introducing a new 'slow' time variable $\tau =t/t_*$,
where the time $t_{*}$ defines the tidal apsidal precession timescale for a small eccentricity $e$ and is given by
\begin{equation}
t_*={{\tilde a}^{13/2}\Omega_*^{-1}\over 15 k_2 q\sqrt{(1+q)}}, 
\label{ev1} 
\end{equation}
where we recall that  $\tilde a=a/R_*,$ 
equation (\ref{betaeq}) thus leads to
\begin{equation}
{d\beta\over d\tau}=\left({\cos \beta \over {\sqrt{(1-e^2)}}}+{1\over \tilde S}\right)\tilde T
(1-e^2)^{3/2}\sin (\beta)\sin(2\hat \varpi),  
\label{ev2} 
\end{equation}
\noindent 
where 
\begin{equation}
\tilde T={3\over 5}(1+q)\gamma_* {e^2(1+{e^2/ 6})\over (1-e^2)^{6}}{\tilde a}^{-3}\sigma^2, 
\label{ev3} 
\end{equation}
and 
\begin{equation}
\tilde S={\tilde I(1+q)\over q}{\tilde a}^{-2}\sigma, 
\label{ev4} 
\end{equation}
where we recall that 
$\sigma = \Omega_r/ n_0$, in addition $\omega_*=\omega_{eq}/\Omega_*,$ ${\tilde I}= I/(M_*R_*^2),$  and\\ $\gamma_{*}=(2\beta_{*}+1)/(2 \omega_{*}^2)$ is a numerical factor 
order of unity.
The dimensionless quantity $\tilde T$ is related to  the ratio of the torque $T_*$ introduced in (\ref{TX})
and the orbital angular momentum and $\tilde S/\sqrt{1-e^2}$   is the ratio of the spin and orbital angular momentum.
In what follows we set $\gamma_*=1$ and adopt 
$\tilde I=0.1$. 

Eq. (\ref{prec}) together with (\ref{ev1})  
leads to the representation of the apsidal precession rate in terms of the dimensionless time, $\tau$,  in the form
\begin{equation}
\frac{{d\hat \varpi}}{d\tau}={d \varpi_{T}\over d\tau}+ {d \varpi_{E}\over d\tau}+
{d \varpi_{R}\over d\tau}+{d \varpi_{NI} \over d\tau},
\label{ev5} 
\end{equation}
where
\begin{equation}
\hspace{-2mm}{d \varpi_{T}\over d\tau}={\left( 1 +3 e^2/2  + e^4/8\right )\over (1-e^2)^5},
\label{ev7} 
\end{equation}
\begin{equation}
\hspace{1.7cm}{d \varpi_{E}\over d\tau}\approx 
4.3\times 10^{-5}\alpha_{E}{(1+q)\over q}{\tilde a^{4}\over (1-e^2)}, 
\label{ev6a} 
\end{equation}
\begin{equation}
\hspace{6mm}{d \varpi_{R}\over d\tau}={{(1+q)}\over 30}{1\over q}{(3\cos^2\beta -1)\over (1-e^2)^2}
\sigma^2,
\label{ev6} 
\end{equation}
\begin{equation}
{\rm and}\hspace{2mm}{d \varpi_{NI} \over d\tau}={1\over 15\tilde I}\left({J\over L}\right){\cos i \cos \beta \over (1-e^2)^{3/2}}\sigma {\tilde a}^{2},
\label{ev8} 
\end{equation}
\begin{align}
\hspace{2mm}{\rm with}\hspace{2mm}\alpha_{E}=\left({M_*\over M_{\odot}}\right)\left({k_2\over 10^{-2}}\right)^{-1}\left({R_*\over R_{\odot}}\right)^{-1}.
\end{align}

The dependence on $\cos i$ can be removed by using the relation for the component 
of the total angular momentum in the direction of ${\bf L}$
\begin{align}
J\cos i = L+S\cos\beta=L(1+(1-e^2)^{-1/2}{\tilde S})\label{JoverLcos}
\end{align}
substituting  the above into (\ref{ev8}) and making use of (\ref{ev4})
we obtain
\begin{align}
{d \varpi_{NI} \over d\tau}={ \sigma {\tilde a}^{2}\cos \beta \over  15{\tilde I}(1-e^2)^{3/2}} +\frac{(1+q)\sigma^2\cos^2\beta}{15q(1-e^2)^2}   .
\label{ev81}
\end{align}

Note that although the Einstein term (\ref{ev6a}) contains a small parameter, it dominates over the tidal contribution 
when either $\tilde a$ is sufficiently large, or $q$ is sufficiently small. Comparing (\ref{ev7}) and (\ref{ev6a}) we find that the Einstein
term dominates over the tidal one provided that   
\begin{equation}
\tilde a > \tilde a_E \approx 12 \alpha_E^{-1/4}\left({q\over 1+q}\right)^{1/4}{\left(1+{3e^2/ 2}+{e^4/ 8}\right)^{1/4} \over (1-e^2)}.   
\label{ev8a} 
\end{equation}

The set of equations (\ref{ev1}) and (\ref{ev5}) also depend on the eccentricity $e$. 
We recall that $J,S,a,$ and accordingly $\sigma$ and  ${\tilde S}$ are constant in non dissipative evolution (see e.g. IP)
The eccentricity can be expressed in terms
of the angle $\beta$ using the first integral derived from the conservation of total angular momentum given by (\ref{integral}), which leads to the relation
\begin{equation}
{\cal{C}}={(1-e^2)\over 2\tilde S}+\sqrt{1-e^2}\cos \beta.
\label{ev9} 
\end{equation}
{where ${\cal{C}}$ is a constant \footnote{Note a misprint in the corresponding equation (116) of IP, the sign
(-) on r.h.s. should be (+). which leads to consistency with (\ref{ev9}).}.
This may be chosen so that the system takes on 
prescribed values $\beta=\beta_0,$
and $e= e_0$ at $\tau=0.$ Thus
\begin{align}
{\cal {C}}={(1-e_0^2)\over 2\tilde S}+\sqrt{1-e_0^2}\cos \beta_0,\label{initC}
\end{align}}

Equation (\ref{ev2}) together with equations (\ref{ev5}-\ref{initC})   form a complete set for determining the evolution as a function of $\tau.$
This is converted to time $t=t_*\tau$  using  (\ref{ev1}). In particular after specifying conserved quantities
and  making use of (\ref{ev7}-\ref{initC}) equations (\ref{ev2}) and (\ref{ev5}) become a pair of 
first order ordinary differential equations for $\beta$ and ${\hat \varpi}$.
These contain, $\tilde a, \sigma, \alpha_E, q $ and $\tilde I$ as fixed parameters.

\subsubsection{Allowed values of $\tilde a$ and $\sigma$}

Here we point out that in what follows,  $\tilde a$ should not be too small, and $\sigma $ should not  be too large. Clearly
the radius of periastron, $r_{p}=(1-e)a$, should be larger than the stellar radius $R_{*}$.
Thus $\tilde a$ should be larger than $1/(1-e)$. Additionally, the  radius  of periastron cannot smaller than tidal disruption 
radius $r_{T}={({M_1/ M_*})}^{1/3}R_*=q^{1/3}R_{*}$,  this being larger than the stellar radius when $q > 1$.
 Combining the requirement that $r_p$ should be larger than both $R_*$ and $r_{T}$ we have
\begin{equation}
\tilde a > \tilde a_{min}={\max (1, q^{1/3}) \over (1-e)}.
\label{lim1} 
\end{equation}
In addition the rotational frequency $\Omega_r$ should be significantly smaller  than $\Omega_{*}=\sqrt{{GM_*/ R_*}^3}$ as when
$\Omega_r \sim \Omega_{*}$ the star experiences rotational break-up. 
Furthermore, for sufficiently large rotation rates the theory 
leading to our evolution equations is not applicable. 
Following \cite{IP2007a} 
we shall assume that $\Omega_{r} < 0.5\Omega_{*}.$ From this  and given that  $\sigma=\Omega_r/n_0 $ we obtain
\begin{equation}
\sigma < \sigma_{max}={{\tilde a}^{3/2}\over 2\sqrt{1+q}}.
\label{lim2} 
\end{equation}
  
\subsubsection{The rate of precession of the longitude of periapsis $d\varPi/dt$}\label{periapsis}

We recall that (\ref{ev8}) as given by (\ref{a15}) is the {contribution to the  rate of advance of the line of apsides measured with respect to the line of nodes
that arises from the precession of the line of nodes itself.}
It can also be written as $-(d\Omega_N/dt)\cos i$, which is $(-)$ the component of the angular velocity associated with the precession of the line of nodes, $d\Omega_N/dt,$ in the direction of the orbital angular momentum.
In order to remove this contribution when either $|\cos\beta|=1$ or in the limit when the magnitude of the spin angular momentum
is negligible compared to the orbital angular momentum,  the longitude of periapsis, $\varPi = \Omega_N + {\hat \varpi}-{\rm \pi}/2$  is often used.
When $|\cos\beta|=1$ the precession is then measured with respect to a line fixed in an inertial frame.
We have $d\varPi/dt= d{\hat \varpi}/ dt+d\Omega_N/dt$.  Thus the transition from $d{\hat \varpi}/dt$  to $d\varPi/dt$    is  obtained if  $\cos i$  is replaced by by  $\cos i -1$ in (\ref{ev8}).
Making use of (\ref{JoverL}) and (\ref{JoverLcos}) one finds that following the above prescription (\ref{ev8}) is modified to become
 \begin{equation}
{d \varpi_{NI} \over d\tau}\rightarrow{1\over 15\tilde I}\frac{(\sqrt{1-S^2/J^2\sin^2\beta}-1)}{(\sqrt{1-S^2/J^2\sin^2\beta}-S\cos\beta/J)}{\sigma {\tilde a}^{2} \cos \beta \over (1-e^2)^{3/2}}.
\label{ev82} 
\end{equation}
Notably, this vanishes in the limit of small $S/J$ which is the expected situation when $q$ is of order unity.
Thus  in this limit  $d\varPi/dt$ is obtained from $d{\hat \varpi}/dt$ by simply omitting  $ d\varpi_{NI}/dt.$
However, no such simplification occurs for small $q$ and it is important to note that ${\hat\varpi}$ rather than $\varPi$
is the significant angle when the evolution of $\beta$ is concerned. Hence, { hereafter we } focus on this.

\section{ Discussion of the evolution equations}
\label{eveq}

\subsection{ A qualitative analysis of the evolution equations under the assumption that variations of $\beta$ are small }
\subsubsection{Determining the dominant form of apsidal precession}\label{s4.1}

The behaviour of our system   depends on the  relative values of
${d \varpi_{T}/ d\tau}$, ${d \varpi_{E}/ d\tau}$, ${d \varpi_{R}/ d\tau}$ and ${d\varpi_{NI} / d\tau}$. 
To estimate 
importance of these terms which contribute to the right hand side of equation (\ref{ev5}), 
we set $(3\cos^2\beta -1)$ and $\cos \beta $ to unity
in equations (\ref{ev6}) and (\ref{ev81}), respectively.  
We then adopt the largest of the two terms on the right hand side of (\ref{ev81}) to make estimates.  In this way we obtain
\begin{align}
&\frac{d \varpi_{NI}}{ d\tau}\sim \max\left(\frac {\sigma {\tilde a}^2}{ 15\tilde I (1-e^2)^{3/2}} , 2\frac{d \varpi_{R}}{ d\tau}\right), \hspace{2mm} {\rm and }\hspace{2mm}\nonumber\\
&\frac{d \varpi_{R}}{ d\tau}\sim \frac{(1+q)\sigma^2}{ 30q (1-e^2)^2}.\label{ev92}
\end{align}
\noindent 
It follows that either 
we have ${d \varpi _{NI}/ d\tau}\sim {d \varpi^{(1)}_{NI}/ d\tau}\equiv {\sigma {\tilde a}^2/ (15\tilde I (1-e^2)^{3/2})}$,
or both inertial and rotational terms have the same order of magnitude. In what follows we call the latter case as
rotational-non-inertial and use ${d \varpi_{RNI}/ d\tau}\sim (~1~+~q~)\sigma^2 / (15q  (1-e^2)^2)$ for our 
estimates below.
 
\subsubsection{ Values of, $\sigma \equiv \Omega_r/n_0,$ separating regimes of tidal and  non inertial precession}

Let us consider the 
situation 
when $\tilde a < \tilde a_{E}$, and,
accordingly, tidal precession is more important than  Einstein precession.
From the condition, ${d \varpi_{RNI}/ d\tau} > {d \varpi_{T}/ d\tau},$ 
we obtain the requirement that  $\sigma > \sigma_1$, where
\begin{equation}
\sigma_1=\sqrt{\frac{{15}q(1+3e^2/2+e^4/8)}{{{(1+q)}}(1-e^2)^{3}}}.
\label{ev11} 
\end{equation}
Similarly, the condition that, 
${d \varpi_{RNI}/ d\tau} > {d\varpi^{(1)}_{NI}/d\tau}$ leads to the requirement  $\sigma > \sigma_2$,
where 
\begin{equation}
\sigma_2={q\over {(1+q)} \tilde I}
(1-e^2)^{1/2}{\tilde a}^{2}.
\label{ev12} 
\end{equation}
In addition, the condition ${d \varpi^{(1)}_{NI}/ d\tau}  > {d \varpi_{T}/ d\tau}$ leads to
$\sigma > \sigma_3$, where
\begin{equation}
\sigma_3={15 \tilde I} 
{(1+3e^2/2+e^4/8)\over (1-e^2)^{7/2}}{\tilde a}^{-2}.
\label{ev13} 
\end{equation} 
\subsubsection{ Values of, $\sigma \equiv \Omega_r/n_0,$ separating  Einstein and non inertial  precession}
When  $\tilde a > \tilde a_{E}$ and Einstein precession is more important than tidal precession,  the condition  
${d \varpi_{RNI}/ d\tau} > {d \varpi_{E}/ d\tau}$ gives $\sigma > \sigma_4$, where
\begin{equation}
\sigma_4={ 2.5}\times 10^{-2}\alpha_{E}^{1/2}
(1-e^2)^{1/2}{\tilde a}^2.
\label{ev13a} 
\end{equation} 
In addition the condition, ${d \varpi^{(1)}_{NI}/ d\tau} > 
{d \varpi_{E}/ d\tau},$  yields $\sigma > \sigma_5$, where
\begin{equation}
\sigma_5=6.7\times 10^{-4} 
\alpha_{E}\tilde I {(1+q)\over q}(1-e^2)^{1/2}{\tilde a}^{2}. 
\label{ev13b} 
\end{equation}

From the above considerations we see that for fixed $\tilde I$, $q$ and $e$  
regions in the  $(\tilde a, \sigma)$ plane can be determined where one of
 ${d \varpi_{T}/ d\tau}$, ${d \varpi_{E}/ d\tau},$
${d \varpi^{1}_{NI} / d\tau},$ or ${d \varpi_{RNI}/ d\tau}$ dominates. 
We denote the largest of  these 
 at a point in the $(\tilde a ,\sigma )$ plane 
 as $\dot \varphi$. 
\subsubsection{Critical curves}
There is a possibility that the contribution of the different terms on the right hand side of (\ref{ev5}) cancel each other in such
a way that we have ${{d\hat \varpi}}/{d\tau}\approx 0$. 
For a given set of values of
$\alpha_E$, $q$ $\tilde I$ and  initial values of $e$ and $\beta$, namely $e_0$ and $\beta_0$, respectively,  
the condition  ${d{\hat \varpi} /d\tau}=0$ 
leads to  an algebraic equation for a curve in the $(\sigma, \tilde  a)$ plane,  referred hereafter to as
a 'critical curve', which may or may not have physical solutions depending on the values of the parameters entering (\ref{ev5}).
An analysis of the evolution of our system near  critical curves is discussed below in Section \ref{S5}.
    
\subsubsection{The variation of $\beta$ in the different regimes of apsidal precession}\label{4.1.2}
Away from a critical curve
a characteristic amplitude of variation of $\beta$ in the course of time,
$\Delta \beta $, can be estimated
as $\Delta \beta \sim {\dot \phi}^{-1} {d\beta / d\tau}$, where ${d\beta/ d\tau}$ is given
by equation (\ref{ev2}). For the purpose of making crude estimates we { 
replace $\cos\beta$ and  $\sin2{\hat \varpi}$ by unity, and
$\sin \beta$ by $\sin\beta_0$,} thus obtaining
\begin{align}
\Delta \beta & \sim { {3\over 5}}{q\over \tilde I \dot \phi}
{e^2(1+{e^2/ 6})\over (1-e^{2})^{9/2}}{\sigma }{\tilde a}^{-1}{ \sin \beta_0} \quad {\rm or} \nonumber\\
\Delta \beta &\sim { {3\over 5}}{(1+q)\over \dot \phi}
{e^2(1+{e^2/ 6})\over (1-e^{2})^{5}}{\sigma^2}{\tilde a}^{-3}{ \sin \beta_0},
\label{ev14} 
\end{align}
depending on whether the second term in brackets in (\ref{ev2}) dominates the first  or vice versa. 
From (\ref{betaeq}) and (\ref{ev2})-(\ref{ev4}), we see that the former case  corresponds
to the orbital angular momentum being larger than the rotational angular momentum, being realised when $\sigma < \sigma_2.$

Substituting estimates of ${d \varpi_{T}/ d\tau}$,  ${d \varpi_{E}/ d\tau}$,  ${d \varpi^{(1)}_{NI}/ d\tau}$, or,
${d \varpi_{RNI}/ d\tau}$ for $\dot\phi$ in (\ref{ev14}),
 we can find a typical amplitude of
variation of $\beta $ in the four regions of the $(\tilde a ,\sigma )$ plane, where these terms respectively dominate. In the first of these regions where ${d \varpi_{T}/ d\tau}$ dominates  equation (\ref{ev14}) becomes\footnote{Note that  the first expression in (\ref{ev15})
  corresponds to the 'standard evolution' considered in IP for which
the apsidal precession is dominated by the tidal term  and the orbital angular momentum is more significant.}
\begin{align}
\hspace{-2mm}\Delta \beta &\sim { {3\over 5}}{q\over \tilde I }
{e^2(1+{e^2/ 6})(1-e^{2})^{1/2}\sigma\over \left( 1 +3 e^2/2  + e^4/8\right ){\tilde a}
 }{ \sin \beta_0}, \hspace{2mm} {\rm or}\hspace{2mm}\nonumber\\
 \Delta \beta &\sim  {3(1+q)e^2(1+e^2/6)\sigma^2\over 5\left (1+3e^2/2+e^4/8 \right){\tilde a}^3}{ \sin \beta_0}
\label{ev15} 
\end{align}
the first alternative applying for $\sigma < \sigma_2$ and the second for $\sigma > \sigma_2.$

Similarly, in the region dominated by  Einstein precession equation (\ref{ev14}) becomes 
\begin{align}
\hspace{-2mm} \Delta \beta& \sim  {1.4\times 10^4}{\alpha_{E}^{-1}q^{2}e^2(1+{e^2/ 6})\sigma\over  \tilde I (1+q) (1-e^2)^{7/2} {\tilde a}^{5}}{ \sin \beta_0}
, \hspace{2mm} {\rm or}\hspace{2mm} \nonumber\\
 \Delta \beta& \sim 1.4\times 10^4{\alpha_{E}^{-1}q e^2(1+e^2/6)\sigma^2\over (1-e^2)^{4}{\tilde a}^7}{ \sin \beta_0},
\label{ev15a} 
\end{align} 
the first alternative applying for $\sigma < \sigma_2$  and the second for $\sigma > \sigma_2.$
Finally, in the region where ${d \varpi_{NI}/ d\tau}$
dominates, which is always the case when $\sigma$ is  sufficiently large,  we find, regardless of the magnitude of $\sigma$ or which term in (\ref{ev92}) dominates, that
\begin{equation}
\Delta \beta \sim {9}q
{e^2(1+{e^2/ 6})\over (1-e^{2})^{3}}{\tilde a}^{-3}{\sin \beta_0}. 
\label{ev17} 
\end{equation}

\subsubsection{ Regimes of evolution as a function of   $\sigma$}

Let us consider how different regimes of evolution arise when $\sigma $ increases and all other
quantities entering the equation for apsidal precession rate are kept fixed.  
Firstly consider the case $\tilde a < \tilde a_E$ and precession due to tides
is more important than Einstein precession.
From equations 
(\ref{ev6}-\ref{ev8}) it follows that when $\sigma $ is sufficiently 
small, that is less than the smaller of $\sigma_1$ and $\sigma_3,$ 
the evolution will be dominated by tidal effects. \\

On the other hand, when $\tilde a > \tilde a_E$ 
and Einstein precession is more important than tidal precession,
when    $\sigma$  is less than the smaller of $\sigma_4$ and $\sigma_5$ the evolution will be dominated by Einstein precession. 
When the evolution is dominated by  either  tidal or Einstein precession, the situation is referred to hereafter  as 
"the standard evolution regime".  When this is not the case we designate the situation
 as 'the rotational regime' for any value of $\tilde a$.

\subsubsection{Estimated change in $\beta$ in the different regimes when precession due to tidal effects is more important
than Einstein precession}\label{tidalest}
Let us consider the case $\tilde a < \tilde a_{E}$ in more detail.  
It is easy to see  from their  definitions that if any two of  the $\sigma_i,$ $ i=1,2,3,$ 
are equal then all of them are.
Thus $\sigma_1(\tilde a_*)=\sigma_2(\tilde a_*)=\sigma_{3}(\tilde a_*)$
for any value $\tilde a_*$ of $\tilde a$ for which this occurs.
In stating this we remark that $\sigma_1$ does not in fact depend on $\tilde a.$ 
Note too that the parameters of the problem should be such that $\tilde a_*$  exceeds 
$\tilde a_{min}=1/(1-e)$ in order  for this quantity to play a role. 

Equating $\sigma_1$ and $\sigma_2$ we get
\begin{equation}
\tilde a \equiv \tilde a_{*}=\left({{15(1 +3 e^2/{2}  + e^4/{8})}{(1+q)}  \over q}\right)^{1/4}{{{\sqrt{\tilde {I}}}}\over (1-e^2)}.
\label{ev17a} 
\end{equation}

\noindent As  $\sigma_1$ is independent of $\tilde a$,
$\sigma_2 \propto {\tilde a}^2$ and  $\sigma_3 \propto {\tilde a}^{-2},$ we see that when 
$\tilde a < \tilde a_*$   we have $\sigma_2 < \sigma_1 < \sigma_3.$
Thus  the evolution is in the standard regime when $\sigma < \sigma_1$ 
and  in the rotational regime when $\sigma > \sigma_1$.
When $\sigma < \sigma_2$ in the standard regime we should use the first expression for $\Delta \beta $ in (\ref{ev15})
otherwise the second is used.  In the rotational regime
(\ref{ev17}) should be used.
 
On the other hand, when $\tilde a > \tilde a_*,$  we have  $\sigma_3 < \sigma_1 < \sigma_2$. 
the evolution  is in the standard regime when $\sigma < \sigma_3.$ As the 
orbital angular momentum is more important than
the rotational angular momentum,  the first expression in (\ref{ev15}) should be used. 
When $\sigma > \sigma_3$ the system is in the rotational regime and (\ref{ev17})  applies.  
Equations (\ref{ev15}-\ref{ev17}) indicate
that $\Delta \beta$ increases with $\sigma$ in the standard regime and does not depend on $\sigma$ in the  rotational regime.

{\subsubsection{Estimated  change in $\beta$ in the different regimes when Einstein precession is more important  than precession driven by tidal effects}\label{Einsteinest}
When $\tilde a > \tilde a_{E}$  from (\ref{ev12}), (\ref{ev13a}) and (\ref{ev13b})  we see that
$\sigma_2$, $\sigma_4$ and $\sigma_5$ have the same dependence on $\tilde a$, being $\propto {\tilde a}^2$ 
Thus  the condition for  the non-inertial regime of evolution ${\sigma_2}/\sigma_5 > 1$ is the same for all $\tilde a > \tilde a_E$. It
becomes a condition for mass ratio, $q,$ to be sufficiently large
\begin{equation}
{q\over (1+q)} > 2.5\cdot 10^{-2}\alpha_E^{1/2}\tilde I. 
\label{ev21} 
\end{equation}

 When this condition is satisfied we have $\tilde a_{*} < \tilde a_{E}$, and 
 the  evolution is in the standard regime when 
$\sigma < \sigma_5$ and is rotationally dominated otherwise. 
Since (\ref{ev21}) implies the orbital angular momentum exceeds the
rotational angular momentum, for standard evolution we  use the first expression in (\ref{ev15a}) for $\Delta \beta.$ 
In the rotationally dominated  case (\ref{ev17}) should be used.

When the inequality (\ref{ev21}) is reversed we obtain  standard evolution when $\sigma < \sigma_4$ and  rotational evolution when  $\sigma > \sigma_4$. 
 When $\sigma < \sigma_4$ and  $\sigma < \sigma_2$, $\Delta \beta $ is determined by the first expression in (\ref{ev15a}) and  when $\sigma_2 <
\sigma  < \sigma_4 $,  $\Delta \beta $ is determined by the second expression in (\ref{ev15a}). Finally, when $\sigma > \sigma_4,$ 
$\Delta \beta $ should be evaluated using (\ref{ev17}).}

\subsubsection{Approximate boundaries of the regimes of evolution in the  $\sigma, \tilde a$ plane}

{It is important to note that in all cases when $\sigma$ is large enough $\Delta \beta $ is determined by (\ref{ev17}).  This  gives
the largest possible value of $\Delta \beta$ for all $\sigma, $  for given  $q$, $e$ and $\tilde a$ provided that a single
term  dominates the apsidal precession rate given by equation (\ref{ev5}). However, as  mentioned above, there could be a situation where different 
terms in (\ref{ev5})  compensate each other and the apsidal precession rate is close to zero. This situation is considered in the next Section. 

When $\sigma < \sigma_2,$
the  orbital angular momentum is larger than the rotational angular momentum.  
Whether  standard  evolution or evolution in the rotational regime takes place
is determined by the relation of, $\sigma,$ to  $\sigma_i,  i = 1, 3, 4, 5,$ according as to  whether, $\tilde a,$ is larger or smaller than
$\tilde a_E$ and $\tilde a_{*}$
and also on whether $q$ is such that the inequality (\ref{ev21}) is satisfied.

When this is satisfied and 
$\tilde a_* < \tilde a_{E},$  so that tidal precession dominates. When $\tilde a < \tilde a_{*}$
the border between the standard and rotational regimes 
is given by $\sigma =\sigma_1.$ 
When  $\tilde a_{*} < \tilde a < \tilde a_{E}$ this border is given by $\sigma = \sigma_3.$ 
When $\tilde a > \tilde a_{E}$ and the inequality (\ref{ev21}) is satisfied
the border between the standard and rotational regime is given by $\sigma = \sigma_5.$ 
When $q$ is such that the inequality (\ref{ev21}) is not satisfied, this border is given by $\sigma= \sigma_4.$

These borders between standard and rotational regimes of 
evolution can be used to construct  curves that separate regions where standard evolution occurs from those where
rotationally dominant  evolution occurs throughout allowed regions the $( {\tilde a}, \sigma)$
plane for specified values of $e$ and  $q.$ These are
illustrated  in Fig. \ref{curven} for
$q=1$, $0.1$, $10^{-2}$ and $10^{-3}.$ For each of these cases
$e=0.5$, and $\tilde I=0.1.$ 
Thus  $\tilde a_{min}=2$ throughout.
When $q=1,$  $\tilde a_{*}=1$ and $\tilde a_{E}=14.5.$ When $q=0.1$,  
$\tilde a_{*}=1.6$ and $\tilde a_{E}=9.5.$ When
$q=10^{-2}$,  $\tilde a_{*}=2.8$ and $\tilde a_{E}=5.5.$ When
$q=10^{-3},$   $\tilde a_{*}=5$ and $\tilde a_{E}=3.$
 Note that when $q=1$ or $0.1,$ $\tilde a_{*} < \tilde a_{min},$ and  that only in the case with
$q=10^{-3}$ the inequality (\ref{ev21}) is not satisfied.   
Finally, we recall that we set  $(3\cos^2\beta -1)$ and $\cos \beta $ to unity
in equations (\ref{ev6}) and (\ref{ev81}) to obtain these borders. Given the form of these
equations, this should provide a reasonable approximation for $|\cos\beta|$ not too small.
Polar orbits with $\beta =\pi/2$ are discussed separately in Section \ref{polar} below.

\section{Evolution near a critical curve on which {${\lowercase {d}}\hat\varpi/ {\lowercase {d}}\tau=0$} }\label{S5}

For a particular set of the parameters entering eq. (\ref{ev5}) ${d\varpi /d\tau}=0$. For a given set of values of
$\alpha_E$, $q$ $\tilde I$ and some initial value of $\beta$, $\beta_0$, with corresponding  initial eccentricity, $e=e_0$ (see (\ref{initC})), 
the condition  ${d\varpi /d\tau}=0$ can be represented as a curve
$ \tilde a=\tilde a_{0}(\sigma)$.  When $\tilde a$ is close to $\tilde a_0$ the rate of apsidal precession is small and variations
of $\beta$ are expected to be much larger than in the general case discussed above. The curve $\tilde a=\tilde a_0(\sigma)$
is referred to hereafter  as a critical curve.   In this Section we analyse possible forms of  critical curves
and the variation of $\beta$ when
$\tilde a$ is close 
to $\tilde a_0(\sigma).$

\subsection{Properties of  critical curves}\label{critc}
Setting ${d{\hat \varpi} /d\tau}=0$ in  (\ref{ev5}) results in biquadratic equation for $\tilde a_{0}$ with the solutions
\begin{equation}
 \tilde a_0=\pm\left\lbrace {1\over 2A}\left(-B\pm \sqrt{B^2-4AC}\right)^{1/2}\right\rbrace^{1/2},
\label{ev22} 
\end{equation}
where
\begin{align}
\hspace{-2mm}A&=\gamma_E{(1+q)\over q(1-e_0^2)}, \hspace{1mm} B={1\over 15\tilde I}{\sigma\cos \beta_0 \over (1-e_0^2)^{3/2}},\hspace{2mm}{\rm and}\nonumber\\
  C&={{(1+3e_0^2/2+e^4_0/8)}\over (1-e_0^2)^5}+
{{(1+q)}({5}\cos^2\beta_0-1)\over 30q(1-e_0^2)^2}\sigma^2, 
\label{ev23} 
\end{align}
with $\gamma_E=4.3\cdot 10^{-5}\alpha_E$. {It is clear that 
 that only solutions of (\ref{ev22}) that are real and positive  
can be physically relevant.}
{When simplifying expressions it is sometimes} convenient to {display} 
the explicit dependence of the quantities $A$, $B,$ and, $4AC,$ on $\beta_0$ and $\sigma$. 
Accordingly, we set
\begin{align}
&B=b\sigma\cos \beta_0,\hspace{2mm} {\rm  and} \hspace{2mm}
 4AC=d_1+d_2\sigma^2({5}\cos^2\beta_0-1),\nonumber
 \end{align}
 \begin{align}
&\hspace{0mm} {\rm where}\hspace{1mm}  b={1\over 15\tilde I}{1\over (1-e_0^2)^{3/2}},\hspace{0mm}
 \hspace{1mm} d_1={4\gamma_E(1+q)\over q}{ {(1+\frac{3e_0^2}{2}+\frac{e_0^4}{8})}\over (1-e_0^2)^6}, \nonumber\\
 &\hspace{2mm}{\rm and}\hspace{2mm} 
 \hspace{0cm}d_2={2\gamma_E\over 15}{(1+q)^2\over q^2}{1\over (1-e_0^2)^3}. \label{defs}
\end{align}
From their definitions it follows that $b$, $d_1$ and $d_2$ are always positive. {In terms of these quantities
 (\ref{ev22}) gives \footnote{The possible unphysical solution with $\tilde a_0$ has been omitted.}}
\begin{align}
 \tilde a_0&=  \left\lbrace {1\over 2A}(-b\cos \beta_0 \sigma\pm \sqrt{d}) \right\rbrace^{1/2},\hspace{3mm} {\rm where} \nonumber\\
 \hspace{3mm}d&=(b^2\cos^2\beta_0+d_2(1-{5}\cos^2\beta_0))\sigma^2-d_1.
\label{ev24} 
\end{align}

\subsection{Prograde rotation}
\label{prog}
From (\ref{ev24}) it is  seen that when $\beta_0 < {\pi/2}$, $-b\cos \beta_0 \sigma < 0$ and
there can only be one branch corresponding to $(+)$ in (\ref{ev24}).  It is also necessary that
  $d  > b^2\cos^2 \beta_0 \sigma^2$ for the expression in the braces in (\ref{ev24}) to be positive. Thus we require that
 $d_2(1-5\cos^2\beta_0)\sigma^2 > d_1.$ 
Accordingly, we 
can have physical solutions of (\ref{ev24}) only when
 \begin{align}
&d_1 < d_2\sigma^2 \hspace{2mm}{\rm and }\hspace{2mm} \beta_0 > \beta_{crit}=\cos^{-1}{ \sqrt{ \frac{1}{{5}} - {d_1\over {5}d_2\sigma^2}}}\equiv \nonumber\\
&\cos^{-1}{\sqrt{\frac{1}{5}-\frac{{6}q(1+3e_0^2/2+e_0^4/8)}{(1+q)\sigma^2(1-e_0^2)^3}}}
\label{ev25} 
\end{align}
{Corresponding to prograde rotation we have  $\beta_{crit} <  \beta_0 < \pi/2$}  
and  there is  only one physical solution corresponding to the  $(+)$ sign alternative of the square root  in (\ref{ev24}).
{Note that the condition, $d_2\sigma^2>d_1,$ that is required for the argument of the square root to be positive,
ensures that the apsidal precession rates due to tides and rotational distortion can balance
for some values of $\beta_0$ and $e_0.$} {  From this condition it follows that 
\begin{equation}
\sigma > \sqrt {{30q (1+3e_0^2/2+e_0^4/8)\over (1+q)(1-e_{0}^{2})^{3}}}.
\label{ev25a} 
\end{equation}
On the other hand, for a fixed $\beta_0$, we can formally make $\sigma $ large enough that the last term in the square root
in (\ref{ev25}) can be neglected. In this case $\beta_{crit}=\cos^{-1}{\sqrt{1\over 5}}$ gives the smallest  possible
value of $\beta_0$. However, available values of $\tilde a_{0}$ and $\sigma$ should also satisfy physical conditions (\ref{lim1}) and
(\ref{lim2}). Therefore, there could be a situation when formally possible solutions of (\ref{ev22}) should be ruled out
as unphysical.
\begin{figure}
\begin{center}
\vspace{-1cm}
\includegraphics[width=8.0cm,height= 9.0cm,angle=0]{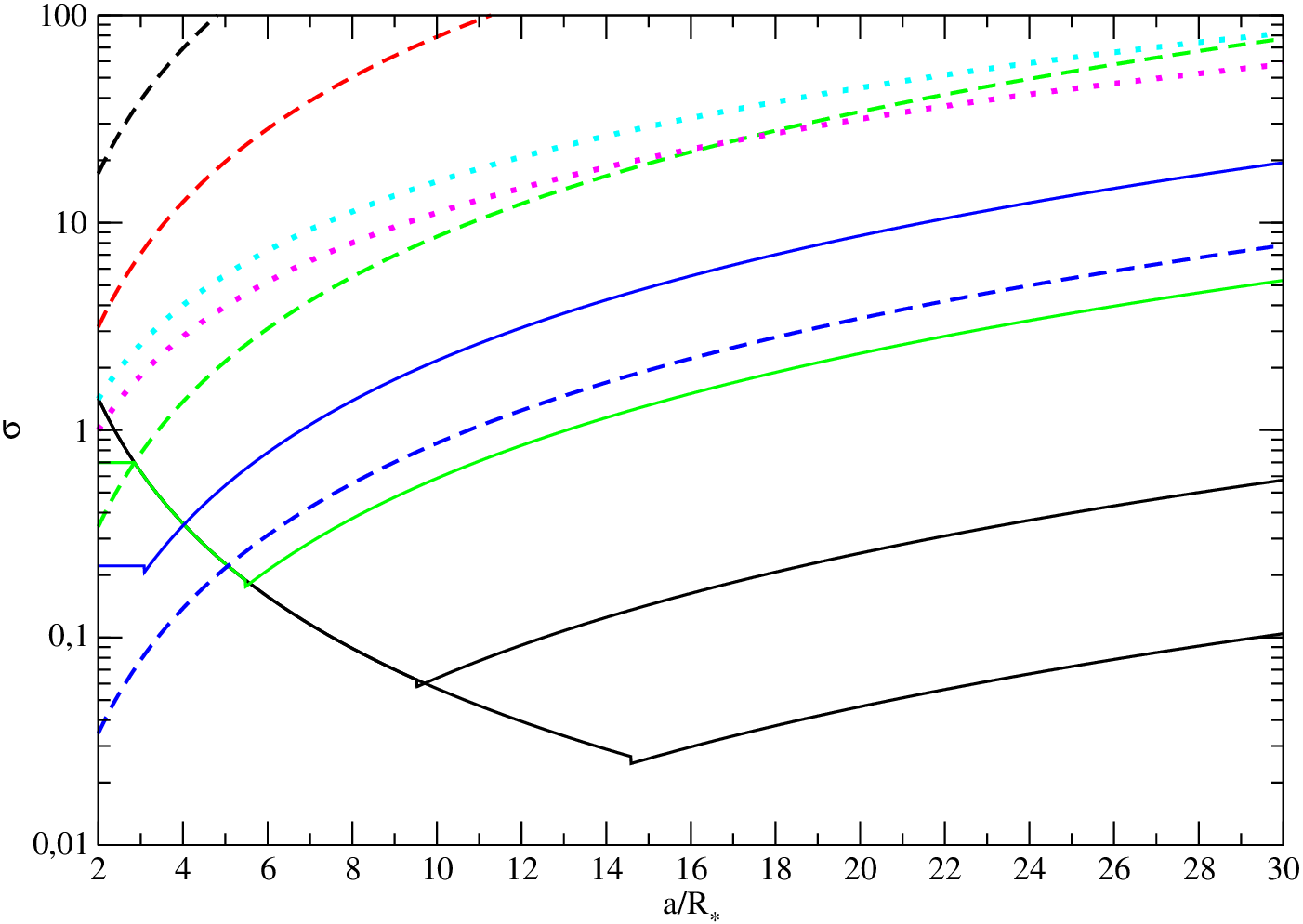}
\vspace{-0cm}
\caption{We show the borders between the standard and rotational regimes represented by solid piecewise continuous curves for
$e=0.5,$  $\tilde I=0.1,$ and various values of $q.$ 
Black, red, green and blue curves correspond to $q=1$, $0.1$, $10^{-2}$ and $10^{-3}$, respectively. 
But note that the first three of these curves have regions of overlap for ${\tilde a} < \sim 10$ on segments
where they are  specified by $\sigma= \sigma_3,$ this quantity being independent of $q.$
\newline The border between the dominance of orbital and rotational angular momenta is defined by the dashed curves. We plot
the maximal value of $\sigma$, $\sigma_{max}$, defined through (\ref{lim2}), magenta and cyan dashed curves correspond to $q=1$ and
to $q=0$, respectively. 
We recall that $\sigma< \sigma_{max}$ to avoid potential rotational disruption.
It is clear  that when $q=1$ or $0.1$ the orbital angular momentum dominates for the whole
range of physically available values of $\sigma$.}
\label{curven}
\end{center}
\end{figure}}

\subsection{Retrograde rotation}
\label{retr}
For  retrograde rotation corresponding to $\beta_0 > {\pi /2}$ 
the quantity  $-b\cos \beta_0 \sigma $ is positive.
In this case, for $d>0,$ if  $d^2 >\sigma^2b^2\cos^2\beta,$  there is   one possible solution  corresponding
to choosing the positive square root
in (\ref{ev24}). On the other hand if $d>0,$ and  $d^2 <\sigma^2b^2\cos^2\beta,$
there are  two possible solutions corresponding to
choosing both the positive and negative square root possibilities
in (\ref{ev24}).

From (\ref{ev24}) the condition that  $d \ge 0$ which is required for the existence of at least one physical solution yields $(b^2\cos^2\beta_0+d_2(1-5\cos^2\beta_0))\sigma^2 \ge d_1.$
When this is satisfied we see that two
 realisable solutions will be present  when
\begin{equation}
\sigma^2d_2(1-5\cos^2 \beta_0)-d_1 < 0.
\label{ev26} 
\end{equation}

Note again that these conditions are only necessary. In addition, $\tilde a_0$ should obviously be larger than $\tilde a_{min}$ and $\sigma $ should be smaller than $\sigma_{max}$. However, in our analysis below
we formally assume that $\tilde a_0$ and $\sigma$ are not constrained by these  physical 
conditions, and instead illustrate them  graphically for a specified value of $\beta_0$.

\subsubsection{The case  $-{1/ \sqrt{5}} < \cos \beta_0 < 0 $}
\label{case1}
The condition that $d > 0$ results in
\begin{equation}
\sigma > \sqrt{d_1\over b^2\cos^2\beta_0+d_2(1-5\cos^2\beta_0)},
\label{ev27} 
\end{equation}
noting that in this case the expression under the square root is always positive. 
From (\ref{ev26}) we  see that two realisable solutions exist when
\begin{equation}
\sigma < \sqrt{d_1\over d_2(1-5\cos^2\beta_0)}.
\label{ev28} 
\end{equation}
As the right hand side of (\ref{ev27}) is always smaller than that of (\ref{ev28}),  there always a region in the  $(\tilde a_0, \sigma)$ plane where two solutions are present subject to the physical 
constraints being met. 

In this region a special role is played by a value of $\sigma=\sigma_d$,
such that  $d$ is zero: $d(\sigma_d)=0$ found by turning the inequality to equality in (\ref{ev27}).
 At this value the branches for each  solution   merge. 
The corresponding dimensionless semi-major axis is $\tilde a_d$. 
When $\sigma$ increases from $\sigma_d$ we have $d>0,$  $\tilde a_0$  is then larger
or smaller than $\tilde a_d$ depending on whether the $(+)$ or $(-)$ sign is adopted for the square root in equation (\ref{ev24}). There are no solutions for $\sigma<\sigma_d$ in this case. 

\onecolumn
\begin{figure}
\begin{center}
\vspace{0cm}
\includegraphics[width=8.0cm,height= 9.0cm,angle=0]{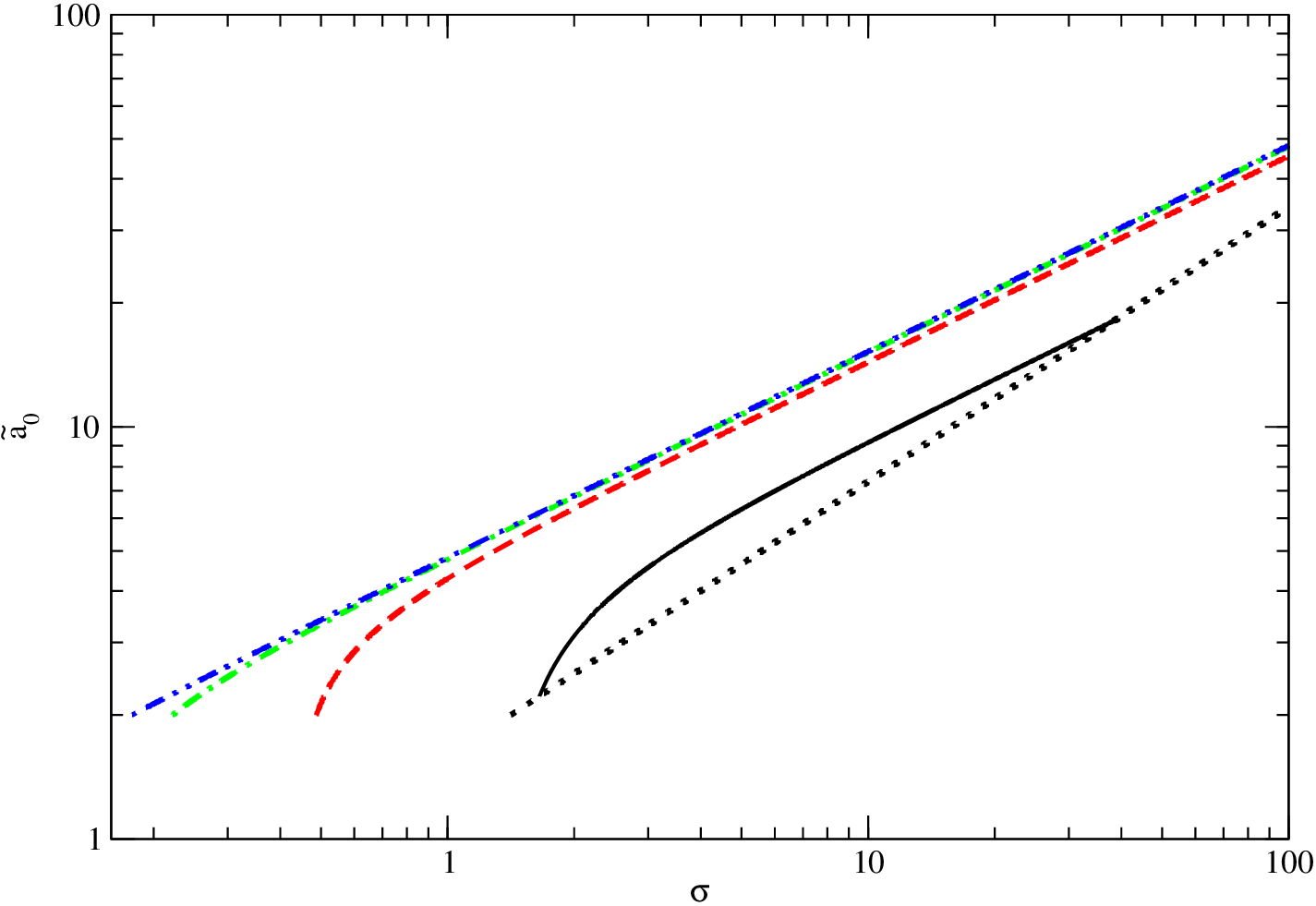}
\hspace{0cm}\includegraphics[width=8cm,height= 9.0cm,angle=0]{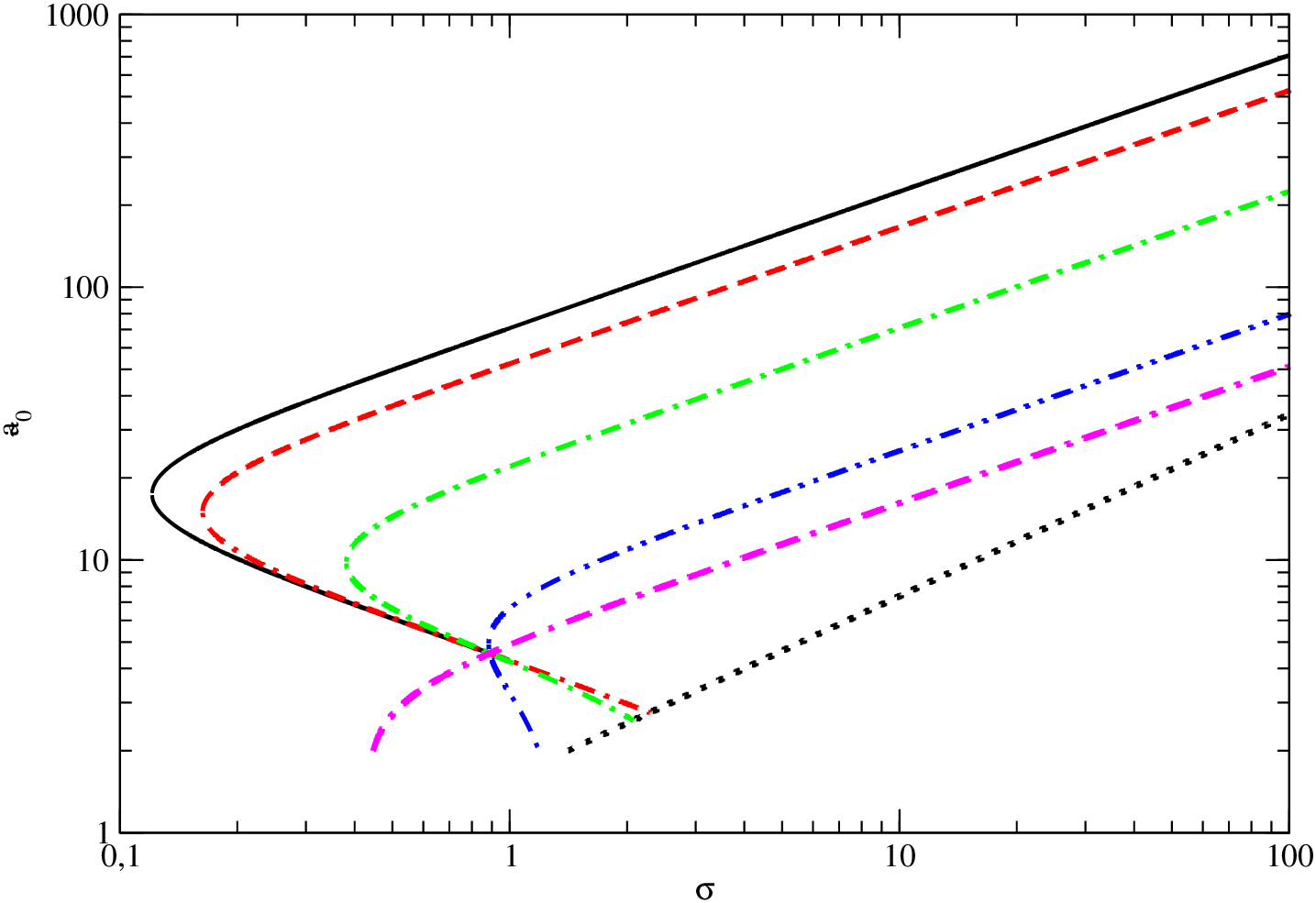}
\caption{The left panel shows critical curves  for the  prograde case with $\beta_{0}= {2\pi / 5}$ in the  
$(\sigma, \tilde a_0)$ plane.
The eccentricity $e_0=0.5$ and  $\tilde I=0.1.$ { Black Solid, red  dashed, green dot dashed and blue 
dot dot dashed curves have $q=10^{-2}$, $10^{-3}$, $10^{-4}$ and
$10^{-5}$, respectively.   The black dotted curve}
 illustrates the condition  (\ref{lim2}) that rotational frequency cannot be too large. 
See the text  for additional  description of particular curves.  
The right panel  is as for the left panel but shows critical curves for  the retrograde case with 
$\beta_{0}={3\pi / 5}$.  In this case
{black solid, red dashed, green dot dashed, blue double dot dashed and  magenta dot double dashed curves are for $q=10$, $1$, $10^{-1}$, $10^{-2}$ and $10^{-3}$ respectively.}}
\label{na_prog}
\end{center}
\end{figure}

\begin{figure}
\begin{center}
\includegraphics[width=8.0cm,height= 9.0cm,angle=0]{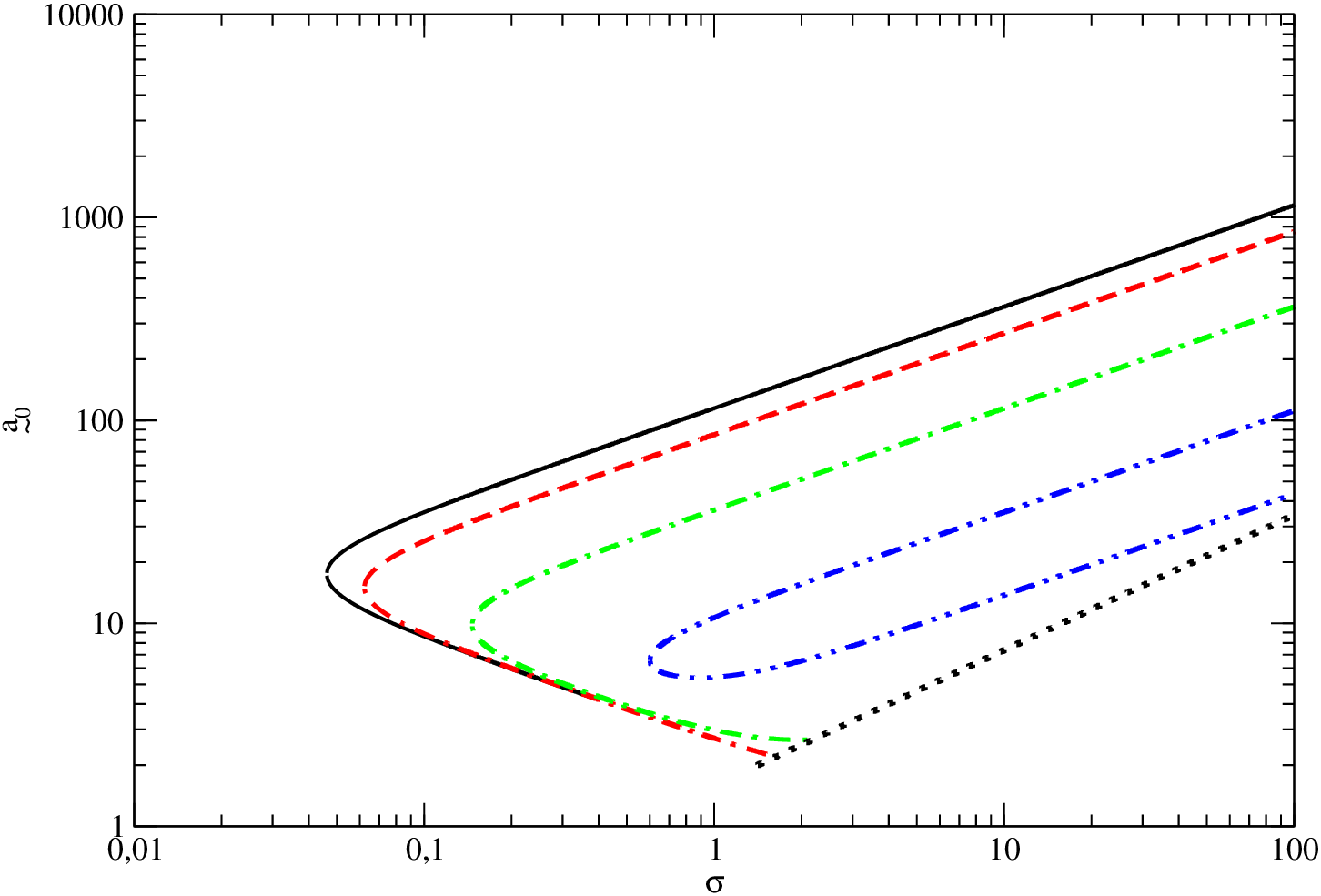}
\hspace{0cm}\includegraphics[width=8cm,height= 9.0cm,angle=0]{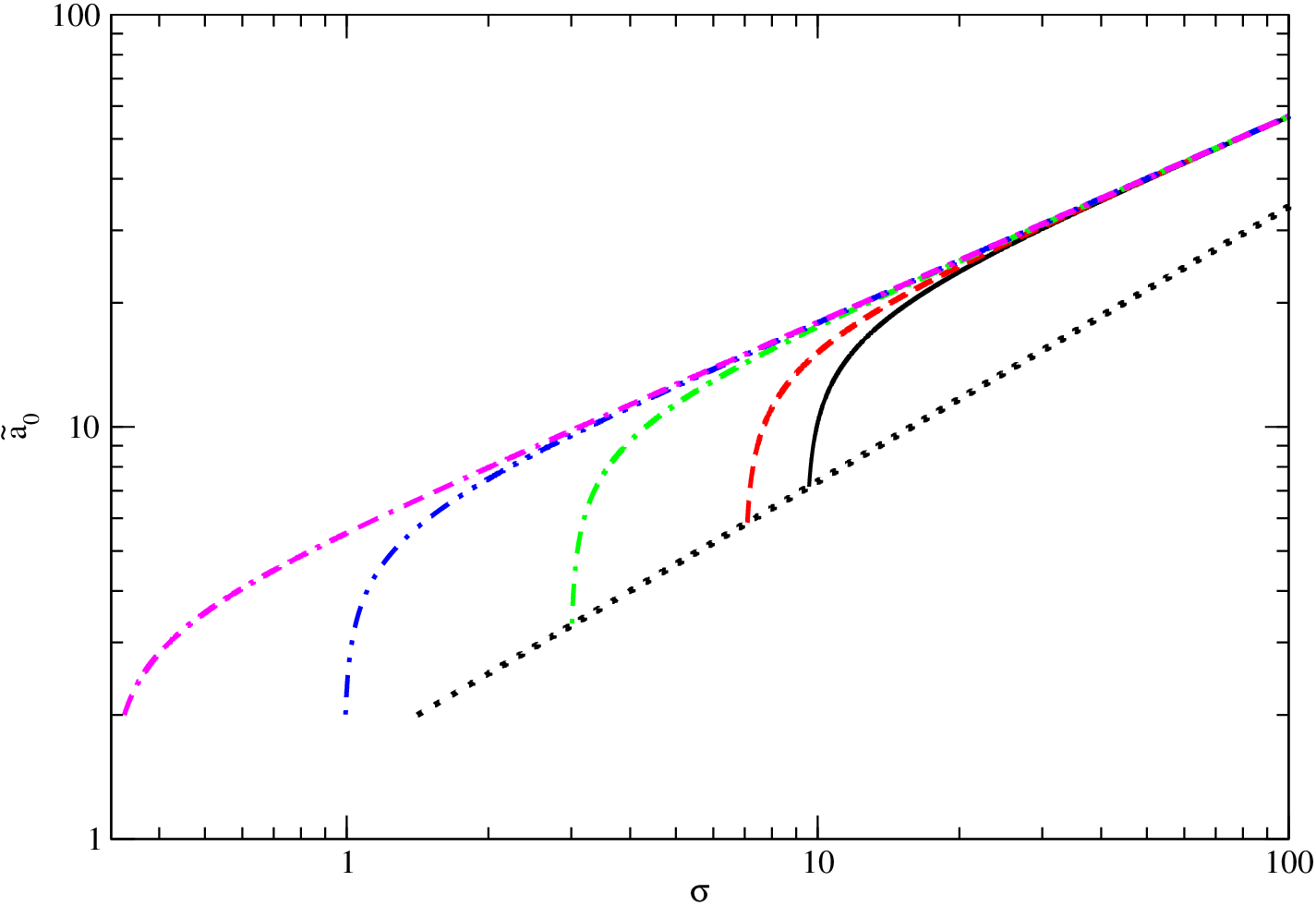}
\caption{The left panel is as in Fig. \ref{na_prog} but for the retrograde case with $\beta_{0}={4\pi / 5}$.
In this case { black solid, red dashed, green dot dashed and blue double dot dashed curves} are for $q=10$, $1$, $10^{-1}$ and $10^{-2}$ respectively.
The right  panel  is as in Fig. \ref{na_prog} but critical curves for polar orbits with $\beta_{0}=\pi/2$ are shown.  In this case
{black solid, red dashed, green dot dashed, blue double dot dashed and magenta dot double dashed curves} are for $q=10$, $1$, $10^{-1}$, $10^{-2}$ and $10^{-3}$respectively.}
\label{na_0_e05}
\end{center}
\end{figure}

\begin{figure}
\begin{center}
\vspace{-1.5cm}
\includegraphics[width=8cm,height= 9.0cm,angle=0]{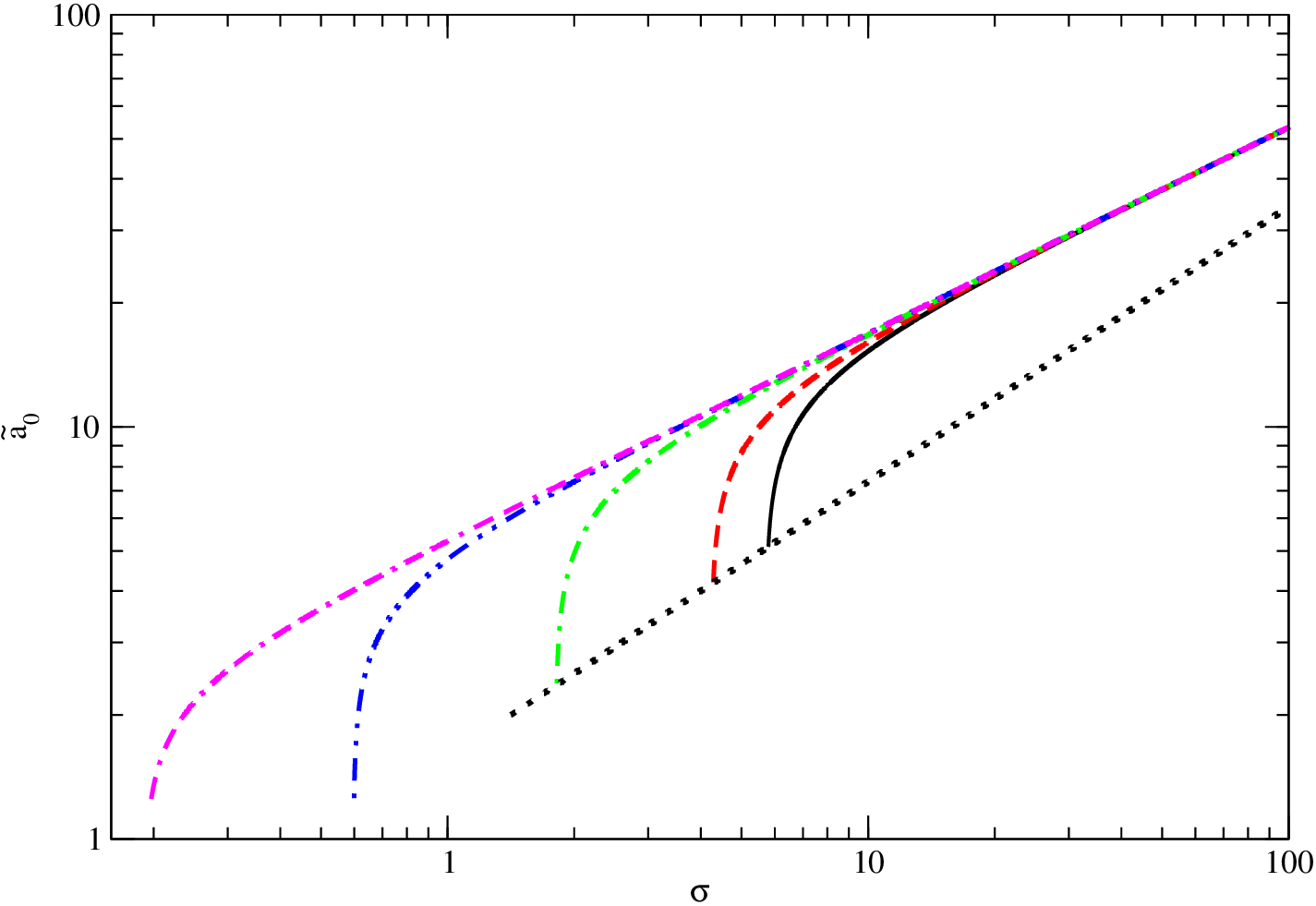}
\hspace{0cm} \includegraphics[width=8.0cm,height= 9.0cm,angle=0]{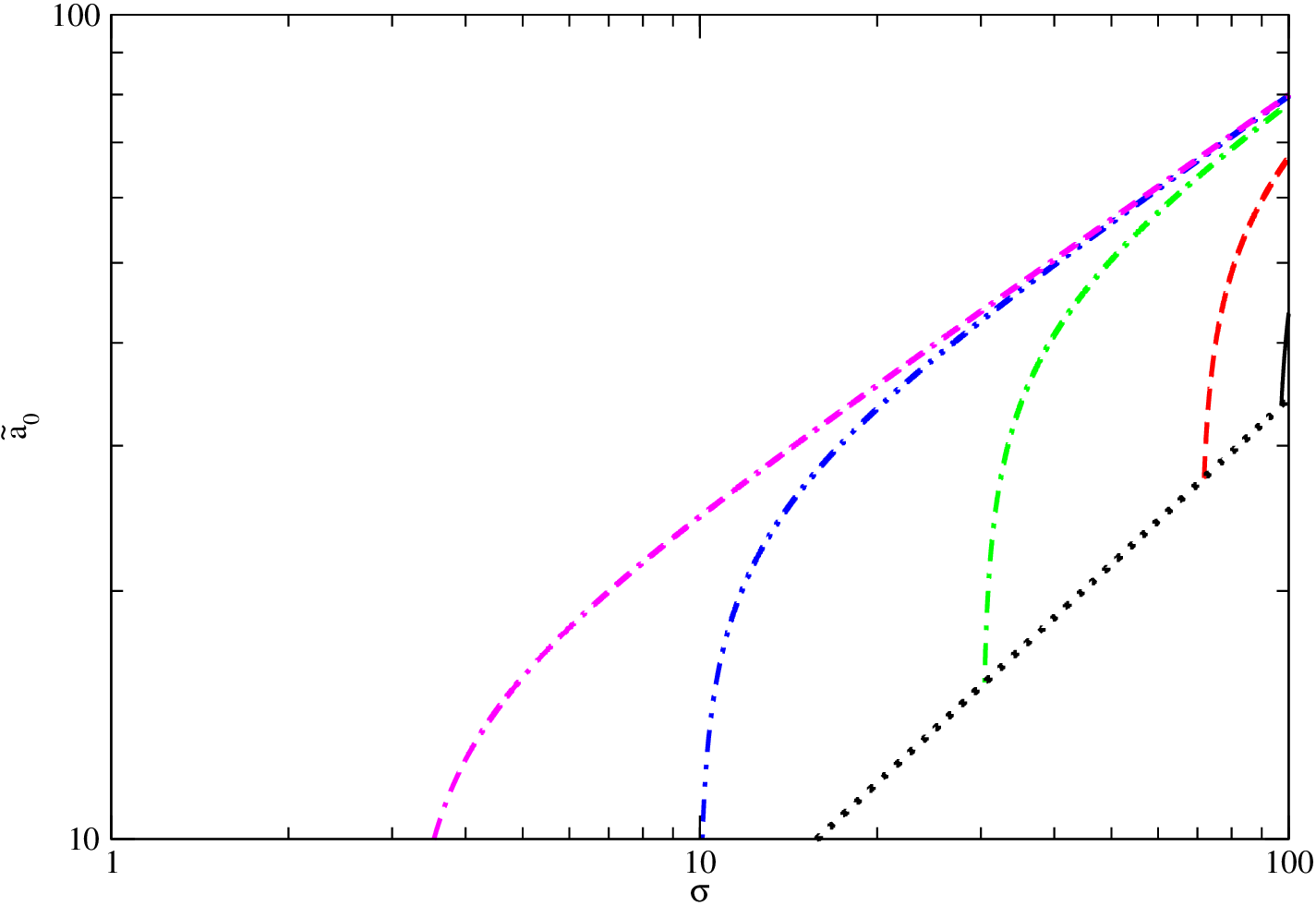}
\caption{As for {the right panel of}  Fig. \ref{na_0_e05}, but with eccentricity $e_0=0.2$ (left panel)
and $e_0=0.9$ (right panel).}
\label{na_0_e02}
\end{center}
\end{figure}

\begin{figure}
\begin{center}
\vspace{2cm}
\hspace{-0.5cm}
\includegraphics[width=8.0cm,height= 9.0cm,angle=0]{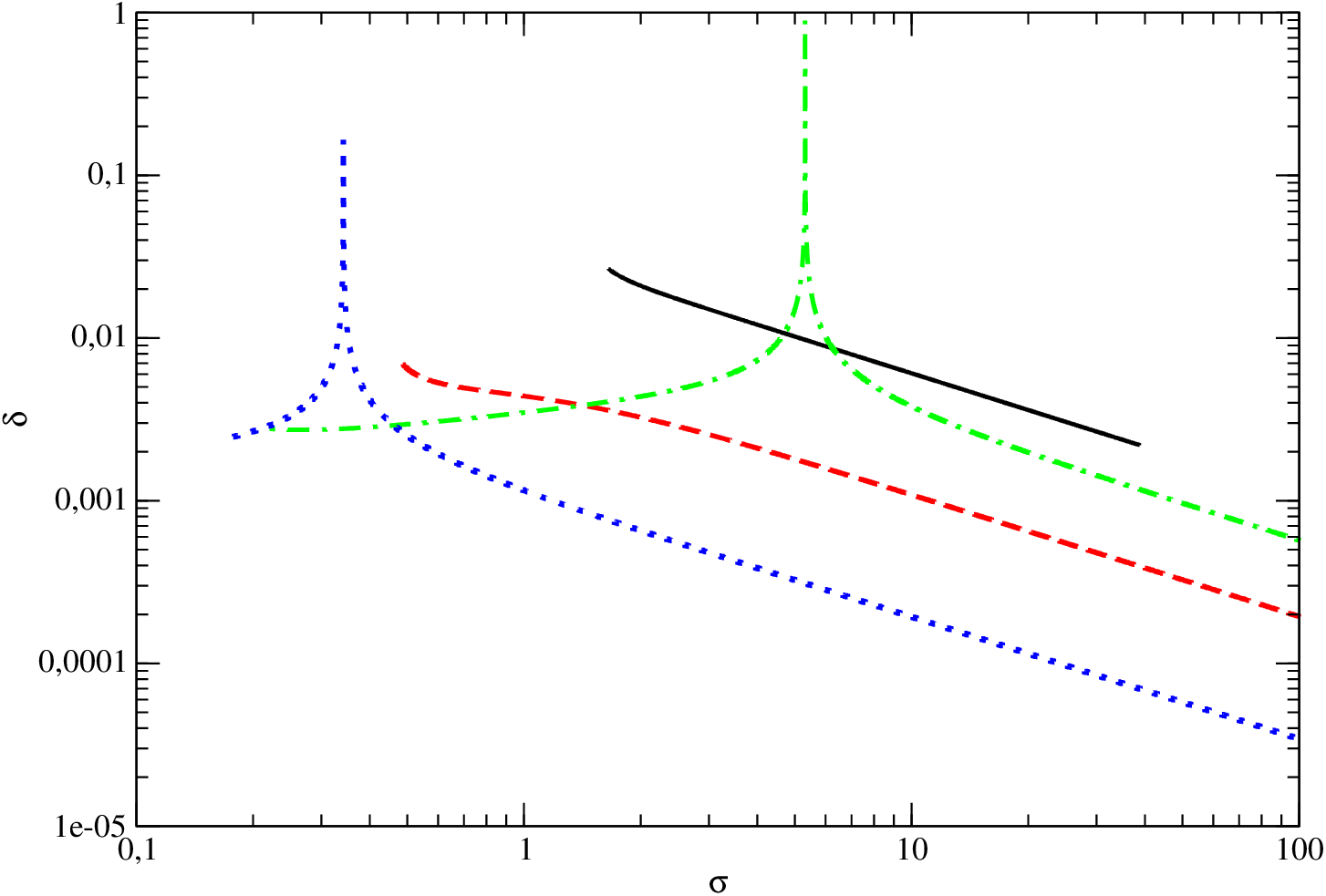}
\hspace{0.5cm}
\includegraphics[width=8.0cm,height= 9.0cm,angle=0]{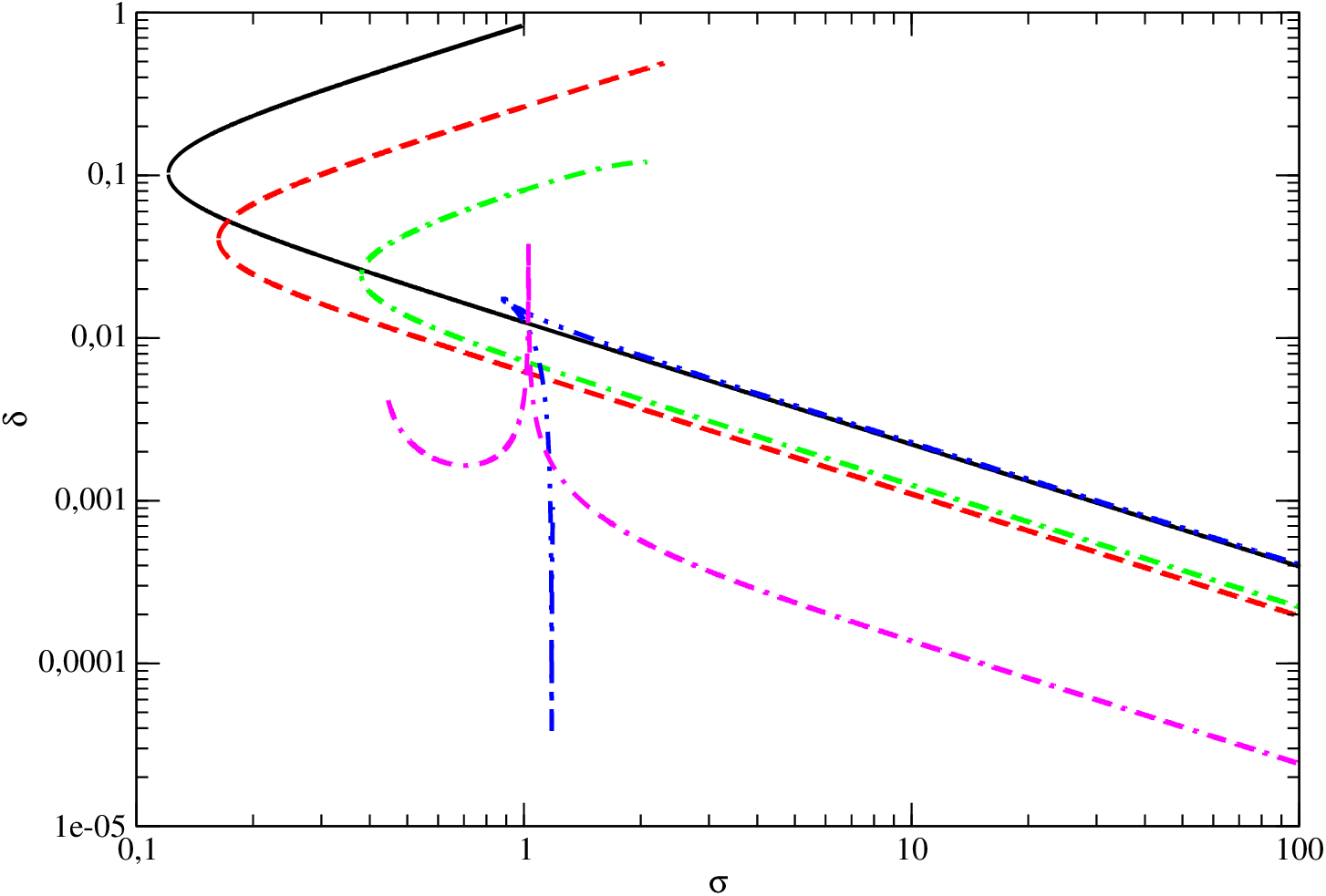}
\caption{{ The characteristic amplitude of the variation of $\beta$ defined in eq. (\ref{libampl1}), and evaluated along the critical curves as a function
of $\sigma$ is shown in the left panel for $\beta_0={2{\rm \pi}/ 5}.$   Curves with different line  styles correspond to values of $q$ in the same way
as for the left panel of Fig. \ref{na_prog}
The same quantity, but for $\beta_0=3{\rm \pi}/5$ is shown in the right panel. In this case the line styles associated with the different curves
are related to $q$ as in the right panel of Fig. \ref{na_prog}. For both panels $e_0=0.5.$ 
{As noted in Section \ref{ampest} the sharp maxima that can be seen are unrealistic owing  to the vanishing of ${\cal D}.$
In practice, as discussed there,  the wings on either side should connect smoothly.
}}}
\label{nd_prog}
\end{center}
\end{figure}

\onecolumn

\subsubsection{The case  $ \cos \beta_0 < -{1 / \sqrt{5}} $}
\label{case2}

In this case the condition that $d$ is positive is again given by (\ref{ev27}), and the expression under 
the square root is  positive for any $\beta_0$  when $b^2 > 5d_2$.  When $b^2 < 5d_2$  it is positive only
when
\begin{equation}
\cos^2\beta_0 \le {d_2\over 5d_2-b^2}.
\label{ev29} 
\end{equation}
The condition (\ref{ev29}) does not constrain possible values of $\beta_0$ provided that the expression on the right hand side is
larger than one. The latter requirement results in
\begin{equation}
{q^{2}\over (1+q)}  > {q_{crit}^{2}\over (1+q_{crit})^{{2}}}={120}\gamma_E {\tilde I}^2. 
\label{ev30} 
\end{equation}
Noting that the smallness of ${\tilde I}$ and  $\gamma_E$  we neglect $q_{crit}$ in the factor $(1+q_{crit})$ in (\ref{ev30})  thus obtaining
\begin{equation}
q_{crit}={7.2}\cdot 10^{-2}\alpha_{E}^{1/2}{\tilde I}. 
\label{ev29a} 
\end{equation}  
In summary, when $q < q_{crit}$ the value of $\cos \beta_0$ should be larger than $-\sqrt{{d_2/( 5d_2-b^2)}}$ for the existence of critical
curves. 
From equation (\ref{ev26}) it is seen that when they do 
exist there always  two solutions  of (\ref{ev24}).

\subsection{Polar orbits}
\label{polar}
Equation (\ref{ev22}), yielding values of ${\tilde a_0}$ on a critical curve, has a solution with a simple form when the stellar rotational axis lies in the orbital plane and, 
accordingly, $\beta_0=\pi/2$. In this case we have from (\ref{ev24}) the single solution
\begin{align}
\tilde a_{0}&=\left({\sigma^2{(1+q)}-30q(1+3e_0^2/2+e_0^4/8) (1-e_0^2)^{-3}\over 30 \gamma_E(1+q)(1-e_0^2)}\right)^{1/4}\nonumber\\
&\approx 5.3\left({\sigma^2{(1+q)}-30q(1+3e_0^2/2+e_0^4/8) (1-e_0^2)^{-3}\over  \alpha_E(1+q)(1-e_0^2)}\right)^{1/4}
\label{ev35} 
\end{align}  
where we use  the definitions of $d_1$ and $d_2$ given in (\ref{defs}), and, obviously, only values { $\tilde a_0 > \tilde a_{min},$ as specified by (\ref{lim1}), should be considered. 
From the condition $\tilde a_{min}=\tilde a_0(\sigma_{min})$ we obtain the smallest allowed value of $\sigma$ to be given by
\begin{equation}
\sigma_{min}=\left ({16(A\tilde a_{min})^4+d_1\over d_2}\right )^{1/2}. 
\label{ev36} 
\end{equation} 
In the same way the largest allowed value of $\tilde a_0$, $\tilde a_{max}$ is obtained by substituting of (\ref{lim2}) in (\ref{ev35}), thus  $\tilde a_{max}=\tilde a_{0}(\sigma_{max})$.}  

\subsection{Graphical representation of critical curves}
\label{figs}

We illustrate realisable   critical curves in the  $(\sigma, \tilde a_0)$ plane  in Figs. \ref{na_prog},   
\ref{na_0_e05}, 
and \ref{na_0_e02} 
for different values of $\beta_0$.  The eccentricity, $e_0$
is taken to be $e_0=0.5$ in Figs. \ref{na_prog}, 
and \ref{na_0_e05}.
In Fig. \ref{na_0_e02} we illustrate critical curves for polar orbits with $\beta_0=\pi/2.$ 
with $e=0.2$  (left panel) and $e=0.9$ (right panel), respectively. 
Curves with different line style are for  different values of $q$ except  for  dotted curves 
which always represent the limiting curve determined  by equation (\ref{lim2}), in which we  have set $q=1.$ 
We remark  that we consider values of $q \le 10$, 
and the difference in  the maximum allowed $ {\tilde a_0}$, $ \tilde a_{lim},$  as determined from (\ref{lim2})  is  a factor  of $\sim 2$ or less. 
In this way   we obtain  ${\tilde a_0} < {\tilde a}_{lim},$ where
$\tilde a_{lim}=(2\sigma)^{2/3}$, which is represented by dotted curves. It is implied, for a given $\sigma$, that only values of ${\tilde a_0} > a_{lim}$ should be taken into account. Additionally, we show only values of ${\tilde a_0}$ larger than $\tilde a_{min}$ given
by equation (\ref{lim1}). 

In the left panel of Fig. \ref{na_prog} we show critical curves on which $\beta_{0}={2\pi / 5}$, which being $< {\pi / 2}$  corresponds to  prograde rotation  as  discussed in Section \ref{prog}. As explained 
there,  there is only one branch of the curves in this case. Also,  only rather small mass ratios are allowed as a result of the condition $\tilde a_0 > \tilde a_{lim}$. 
{Black solid, red dashed, green dot dashed and blue dot dot dashed curves} are for $q=10^{-2}$, $10^{-3}$, $10^{-4}$ and
$10^{-5}$, respectively. As seen from these plots all curves  are such that ${\tilde a_0}$ grows monotonically
 with $\sigma$ and, for a given $\sigma$ larger values of $\tilde a_0$ correspond to smaller mass ratios.

In the right panel of  Fig. \ref{na_prog} we illustrate critical curves with $\beta_{0}={3\pi / 5}$,  which has retrograde rotation 
 and  satisfies $ -{1/ \sqrt 5} < \cos (\beta_{0}) < 0$ as  discussed in Section \ref{case1}. 
{Black solid, red dashed, green dot dashed, blue double dot dashed and magenta dot double dashed curves} are for $q=10$, $1$, $10^{-1}$, $10^{-2}$ and $10^{-3}$, respectively.
As seen from these plots,  the situation is quite different from the previous case.
Apart from the case with $q=10^{-3}$ there are two branches merging at $\sigma=\sigma_d$, which is the smallest  value of $\sigma $ that can be realised on a critical curve with prescribed $q$, $e_0$ and $\beta_0$.
Also, contrary to the previous case values of $\tilde a_0$ for a given $\sigma,$  belonging to the upper branch, are larger for larger values of $q$, values of $\tilde a_0$ corresponding to upper (lower) branch increasing  (decreasing) with $\sigma$.

In the left panel of Fig. \ref{na_0_e05} we illustrate critical curves for
 the retrograde case with $\beta_{0}={4\pi / 5}$. In this case  $\cos (\beta_{0}) < -{1/ \sqrt 5}$ a situation that is discussed in Section \ref{case2}. 
{Black solid, red dashed, green dot dashed and blue double dot dashed curves are} for $q=10$, $1$, $10^{-1}$ and $10^{-2}$, respectively. 
This situation is similar to that previous retrograde case but the curve for  $q=10^{-3}$ is absent. This is because $\sigma_d$ is larger than the largest value of $\sigma,$ namely
$\sigma=100$ shown. Since larger values of $\sigma,$ for which tides are significant,  are unlikely to be realised in an astrophysical context,
we conclude that when  $\cos (\beta_{0}) < -{1/ \sqrt 5}$ and the  mass ratio is sufficiently small,  finding  a system evolving close to a critical curve is unlikely.

Critical curves for polar orbits with $\beta_{0}={\pi / 2}$ with $e_0=0.5$ are shown in  the right panel of Fig. \ref{na_0_e05}.
In addition, critical curves for $\beta_0=\pi/2$ but with $e_0=0.2$ and $e_0=0.9$ are illustrated in the left and right panels of
Fig. \ref{na_0_e02},
 respectively.  {Black solid, red dashed, green dot dashed, blue double dot dashed and magenta  dot double dashed curves} are for $q=10$, $1$, $10^{-1}$, $10^{-2}$ and $10^{-3}$,
respectively. The case $e_0=0.5$ shown in the right panel of Fig. \ref{na_0_e05} can be compared to the previous cases which all have the same value of $e_0$. 
As for the prograde case there is only one branch of a critical curve for a given $q$, but they exist for larger values of $q$   at sufficiently large values of $\sigma$. 
In addition, curves with small mass ratios reach smaller values of $\sigma.$  In the
opposite limit of large $\sigma$ all curves have the same asymptote. As seen from  Fig. \ref{na_0_e02} 
when $e_0$ is smaller (larger) the range of allowed $\sigma$ is shifted towards smaller (larger) values.
Note, however, that in case of  large eccentricity it is more reasonable to compare the rotational frequency with a typical periastron passage frequency, which scales as $ {n_{0}/ (1-e)^{3/2}}$.

{\subsection{The condition $d\varPi/dt=0$ and its limit when the ratio of spin angular momentum to orbital angular momentum is small}
 To obtain $d\varPi/dt$ we equate it to the right hand side of  equation (\ref{ev5}) making use of equations (\ref{ev7})-(\ref{ev6}) for $d\varpi_T/dt, d\varpi_E/dt,$
 and $d\varpi_R/dt$ respectively, and equation (\ref{ev82}) to specify $d\varpi_{NI}/dt.$
 Although the angle $\varPi$ is not directly involved in the evolution of $\beta,$ it may be of interest in the context of observations of apsidal motion
 and when this reverses direction,  which happens when  $d\varPi/dt$ passes through zero.
 When this happens for some $\beta_0,$ and $e_0$ can be determined in the same way as critical curves. To do this, from the discussion in Section \ref{periapsis}
it follows that we should make the replacements
 \begin {align}
&\tilde I \rightarrow \tilde I \frac{(\sqrt{1-S^2/J^2\sin^2\beta}-S\cos\beta/J)}{\sqrt{1-S^2/J^2\sin^2\beta}-1}, \hspace{1mm} {\rm and}\hspace{1mm}\nonumber\\
&5\cos^2\beta\rightarrow
3\cos^2\beta. 
\label{ev82a}
  \end{align}
  In particular this formulation is most useful in the limit $S/J \rightarrow 0$ in which case $B$ in equation (\ref{ev23}) $\rightarrow~0,$  and as a consequence $d\varPi/dt$ passes through zero
  when
  \begin{equation}
{\gamma_E(1+q)\tilde a_0^4\over q(1-e_0^2)}  + {{(1+\frac{3e_0^2}{2}+\frac{e^4_0}{8})}\over (1-e_0^2)^5}+
{{(1+q)}({3}\cos^2\beta_0-1)\sigma^2\over 30q(1-e_0^2)^2}=0.
\label{evhh} 
\end{equation}
 In the case of  polar orbits with $\beta_{0}=\pi/2$ this clearly yields the same critical curve  condition given in 
Section \ref{polar}.
 We also note that as the polar orbit is the most favourable for reversing the sign of $d\varPi/dt,$
 equation (\ref{ev35}) gives an upper bound value  on the values of of  $\tilde a_0,$ for a given $\sigma$, for which this is possible. }

{\subsection{Relationship to fixed points and the evolution of $\beta$}
Critical curves are such that on them $d{\hat \varpi}/d\tau=0.$ As the apsidal precession rate does not
depend on ${\hat \varpi}$ this is not required to define critical curves.
However, if we insist that $d\beta/d\tau =0$ in addition, we define a fixed point.
For general $\beta_0$ this requires ${\hat \varpi}= 0, \pi/2$ or $3 \pi/2$ (see eq. (\ref{betaeq}). 
When ${\hat \varpi}$ takes on one of these values, $\beta$ remains fixed at the value $\beta_0.$
However, if a different value is specified then $\beta$ will vary with time displaying an oscillatory motion.
From (\ref{betaeq})  when  $T_*$ or $e$ is small the amplitude of this motion will be small.
But the changes in $\beta$ will exceed those found well away from critical curves as described 
in Section \ref{4.1.2}.  This will be discussed further below.}

\section{The evolution equations  and the  behaviour of  solutions in the neighbourhood of a critical curve}\label{S6}
\label{evolution}

Let us assume that at the moment of time $t=0,$ $\beta=\beta_0, e = e_0,$ and the solution crosses the critical curve. At this time,
by definition, ${d\hat \varpi}/d\tau=0$ and we have from equations (\ref{ev5}-\ref{ev81})
\begin{equation}
{d \varpi_{T}\over d\tau}+ {d \varpi_{E}\over d\tau}+{(1+q) (5\cos^2\beta_0 -1)\sigma^2\over 30q (1-e_0^2)^2}
+{ \sigma {\tilde a_0}^{2} \cos \beta_0 \over 15\tilde I (1-e_0^2)^{3/2}}=0,
\label{crit1} 
\end{equation}
{When the system evolves with time, the eccentricity changes.
Using the expression of conservation of angular momentum given by $(\ref{ev9})$ we can relate $e$ to $\beta,\beta_0,$ and $e_0.$
With the help of (\ref{initC}) this gives
\begin{align}
e^2-e^2_0&=\frac{2{\tilde S}\sqrt{(1-e_0^2)}(\cos\beta -\cos\beta_0)}{1+2{\tilde S}\cos\beta/(\sqrt{1-e_0^2}+\sqrt{1-e^2})}
\sim\nonumber\\ 
& -\frac{2{\tilde S}\sqrt{(1-e_0^2)}\sin\beta_0(\beta -\beta_0)}{1+{\tilde S}\cos\beta_0 (1-e_0^2)^{-1/2}},
\label{eccb}
\end{align}
where the approximation on the right applies when $\beta$ is close to $\beta_0.$}

{Regarding $d\hat\varpi/d\tau$ as a function of $\beta$ and $e^2,$ as $\beta_0$ and $e_0^2$ correspond to a critical curve, we may write
\begin{align}
\frac{d\hat\varpi}{d\tau}\bigg|_{\beta,e^2}= \frac{d\hat\varpi}{d\tau}\bigg|_{\beta,e^2}- 
\frac{d\hat\varpi}{d\tau}\bigg|_{\beta_0,e_0^2},\label{crit2}
\end{align}
{ where by $d\hat\varpi/d\tau |_{\beta_0,e_0^2}$ we mean the left hand side   of (\ref{crit1}).  This   is   $d\hat\varpi/d\tau$ evaluated for $\beta=\beta_0,$
and $e=e_0$
and of course it is equal to zero. 
Formal subtraction of this expression in (\ref{crit2})  suggests the usefulness of  a first order Taylor expansion. 
This procedure is especially useful in the situation where  we have small changes in $\beta$ 
and the orbital angular momentum is approximately conserved and we have $e \approx e_0$.}  

Equation (\ref{crit2}) together with the equation for $d\beta/d\tau$  
given by equations (\ref{ev2}) and (\ref{ev3}) govern the evolution of the system.
Dividing the second by the first, and then making use of (\ref{eccb}) where necessary, leads to an equation of the generic form
\begin{align}
\hspace{-0cm}{\cal F}(\beta)\frac{d\beta}{d\hat\varpi}=\sin2\hat\varpi, \hspace{2mm}{\rm for}\hspace{2mm}{\rm an}\hspace{2mm}{\rm appropriate}\hspace{2mm}{\rm form}
\hspace{2mm}{\rm of}\hspace{2mm}{\cal F}(\beta).\label{ge1}
\end{align}
This yields on integration
\begin{align}
{\cal G}(\beta)=\cos2\hat\varpi_0- 2\int^{\beta}_{\beta_0}{\cal F}(\beta)d\beta=\cos2\hat\varpi,\label{ge2}
\end{align}
where  $\hat\varpi_0$ is the initial value of $\hat\varpi$ corresponding to $\beta_0.$
Then from (\ref{ge1}) we obtain
\begin{align}
\frac{dy}{d\tau}= \pm\sqrt{1-y^2}{\cal F}^{-1}\sqrt{1-{\cal G}^2}.\hspace{2mm}{\rm where}\hspace{2mm} y=\sin\beta   \label{ge3}
\end{align}
From this it is expected that $y$ is a periodic function of $\tau$
oscillating between positive values such that one of the square roots vanishes.
Given such a periodic solution, there is the possibility that $\hat\varpi$  librates over a restricted domain if  
\begin{align}
\oint_{period} \frac{d\hat\varpi}{d\tau}\bigg|_{\beta,e^2} d\tau=0, 
\end {align}
\footnote{{In this case $\hat\varpi/n$
will librate over a restricted domain provided the integer  $n$ is large enough.}} or ultimately exploring all of $(0,2{\rm \pi})$ otherwise.
Although (\ref{ge3}) 
is soluble 
by quadratures, the integral is not expressible in terms of known functions.
Accordingly, we limit studies to special cases to illustrate these generic features.
\subsection{The evolution when the variation of $\beta$ is small}
This would be expected to occur for example when ${\tilde a}$ is large.
In such a case case $\delta=\beta-\beta_0$ and $e^2-e_0^2$ are small such that we may perform a first order Taylor expansion of
the right hand side of (\ref{crit2}).
This gives
\begin{align}
&\frac{d\hat\varpi}{d\tau}\bigg|_{\beta,e^2}=\left(\frac{\partial  (d\hat\varpi/d\tau)}{\partial \beta}\bigg|_{\beta_0,e_0^2}
-\right.\nonumber\\
&\left.\frac{\partial  (d\hat\varpi/d\tau)}{\partial e^2}\bigg|_{\beta_0,e_0^2}\frac{2{\tilde S}\sqrt{(1-e_0^2)}\sin\beta_0}{(1+{\tilde S}\cos\beta_0 (1-e_0^2)^{-1/2})}\right)\delta
=-{\cal D}\delta,
\label{crit3}
\end{align}
where we have made use of (\ref{eccb}).}

\onecolumn

\begin{figure}
\begin{center}
\hspace{-0.5cm}
\includegraphics[width=8.0cm,height= 7.0cm,angle=0]{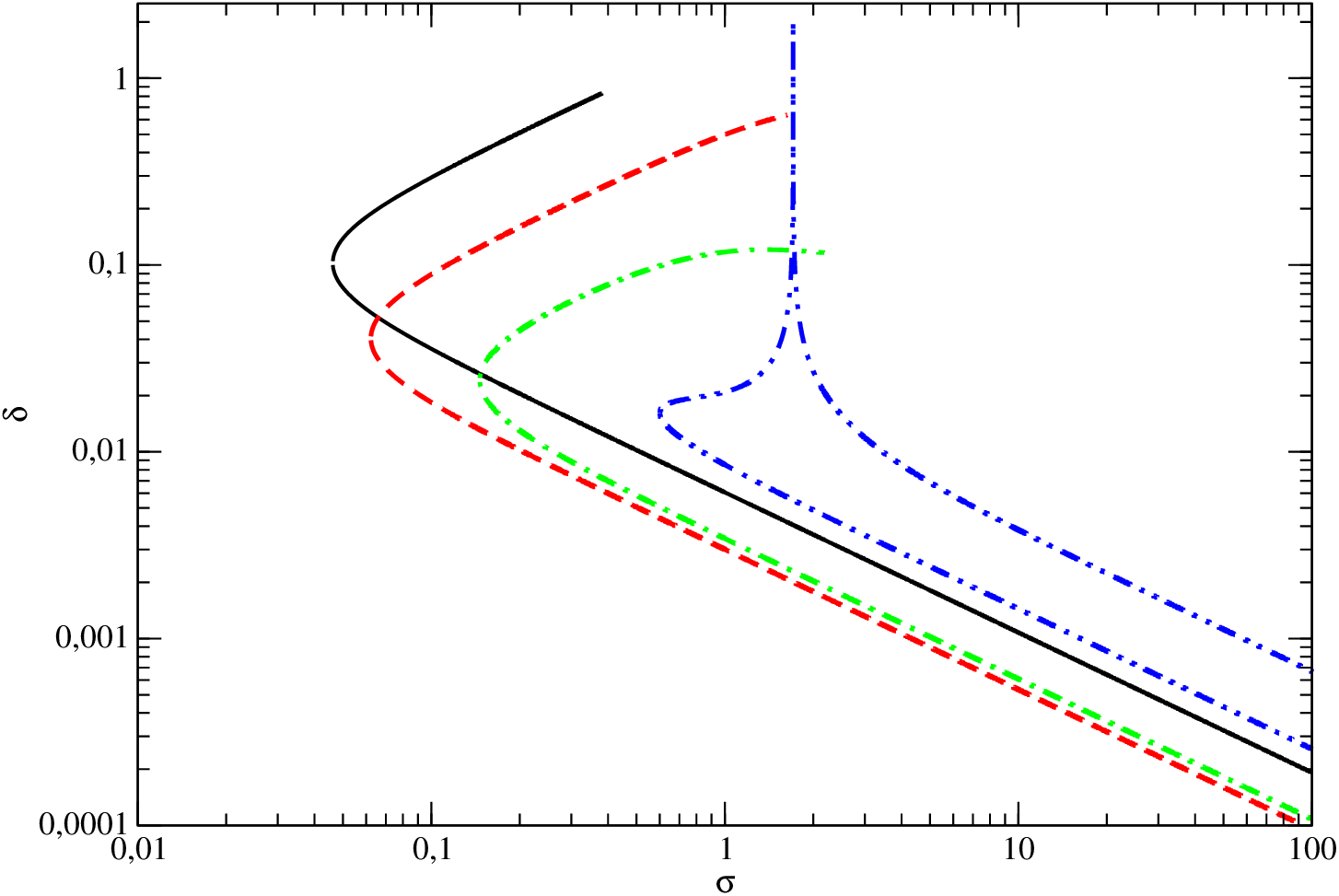}
\hspace{0.5cm}\includegraphics[width=8.0cm,height= 7.0cm,angle=0]{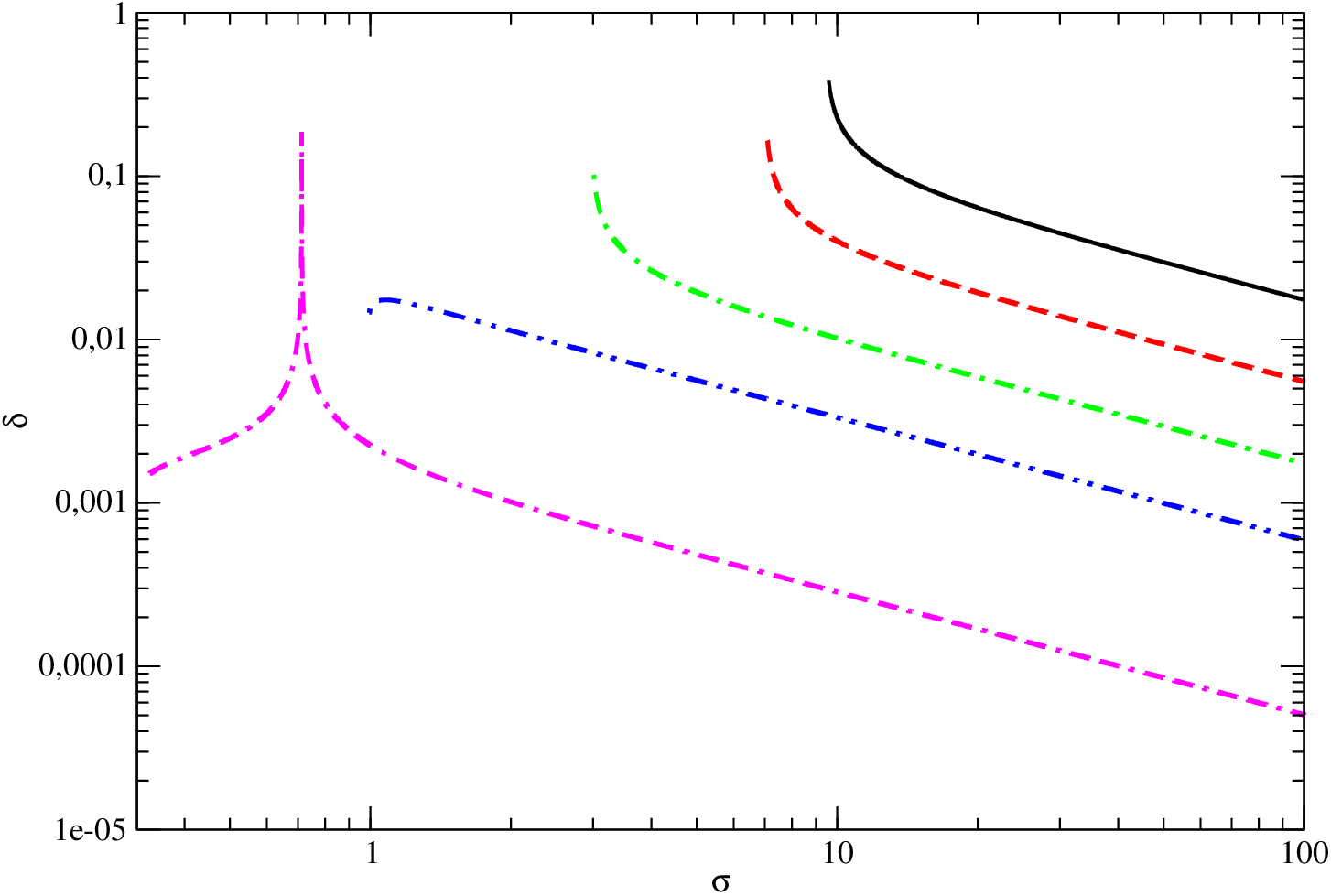}
\caption{{ The left panel is as for  Fig. \ref{nd_prog}, but for  $\beta_0={4{\rm \pi } /5}$. 
In this case the line styles associated with the different curves
are related to $q$ as in the left panel of Fig. \ref{na_0_e05}. The same quantity but for case with $\beta_0= {\rm \pi}/2$
is illustrated in the right panel.  The line styles associated with the different curves are as for the right panel of Fig. \ref{na_0_e05}.
For both panels $e_0=0.5.$}}
\label{nd_retr2}
\end{center}
\end{figure}

\begin{figure}
\begin{center}
\vspace{3cm}
\includegraphics[width=8.0cm,height= 7.0cm,angle=0]{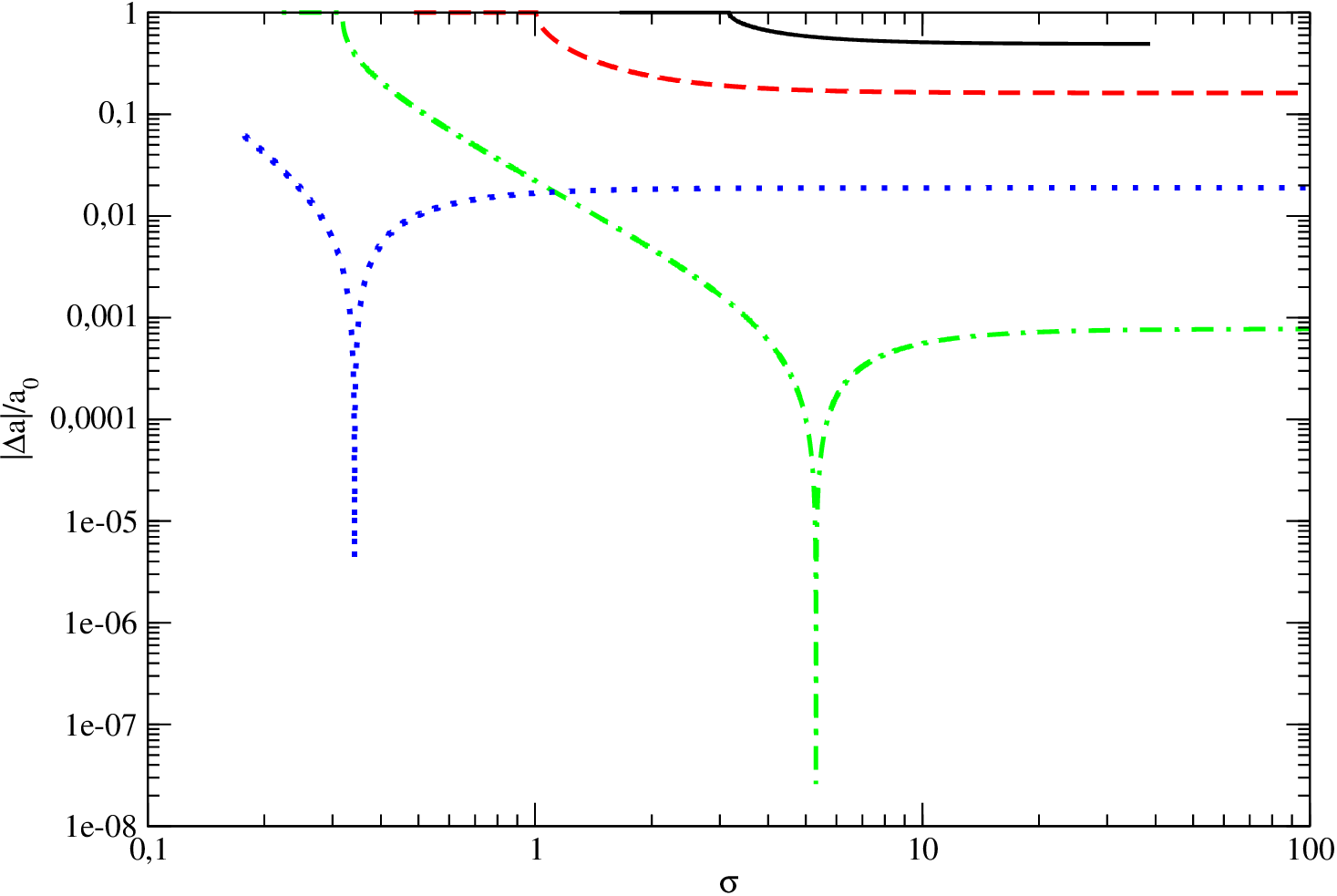}
\vspace{1cm}
\caption{ The quantity  $|(\tilde a_{{\cal D}}(\sigma) -\tilde a_{0}(\sigma))/{\tilde a}_0(\sigma)|$ is plotted as a function of $\sigma.$
This quantity is evaluated for the critical curves illustrated in Fig. \ref{na_prog} for $\beta_0= 2{\rm \pi}/5.$ 
The line styles used are the same as  those of  Fig. \ref{na_prog} such that curves with the same line style correspond to each other.}
\label{ndel_prog}
\end{center}
\end{figure}

\begin{figure}
\begin{center}
\hspace{-1.0cm}
\includegraphics[width=8.0cm,height= 9.0cm,angle=0]{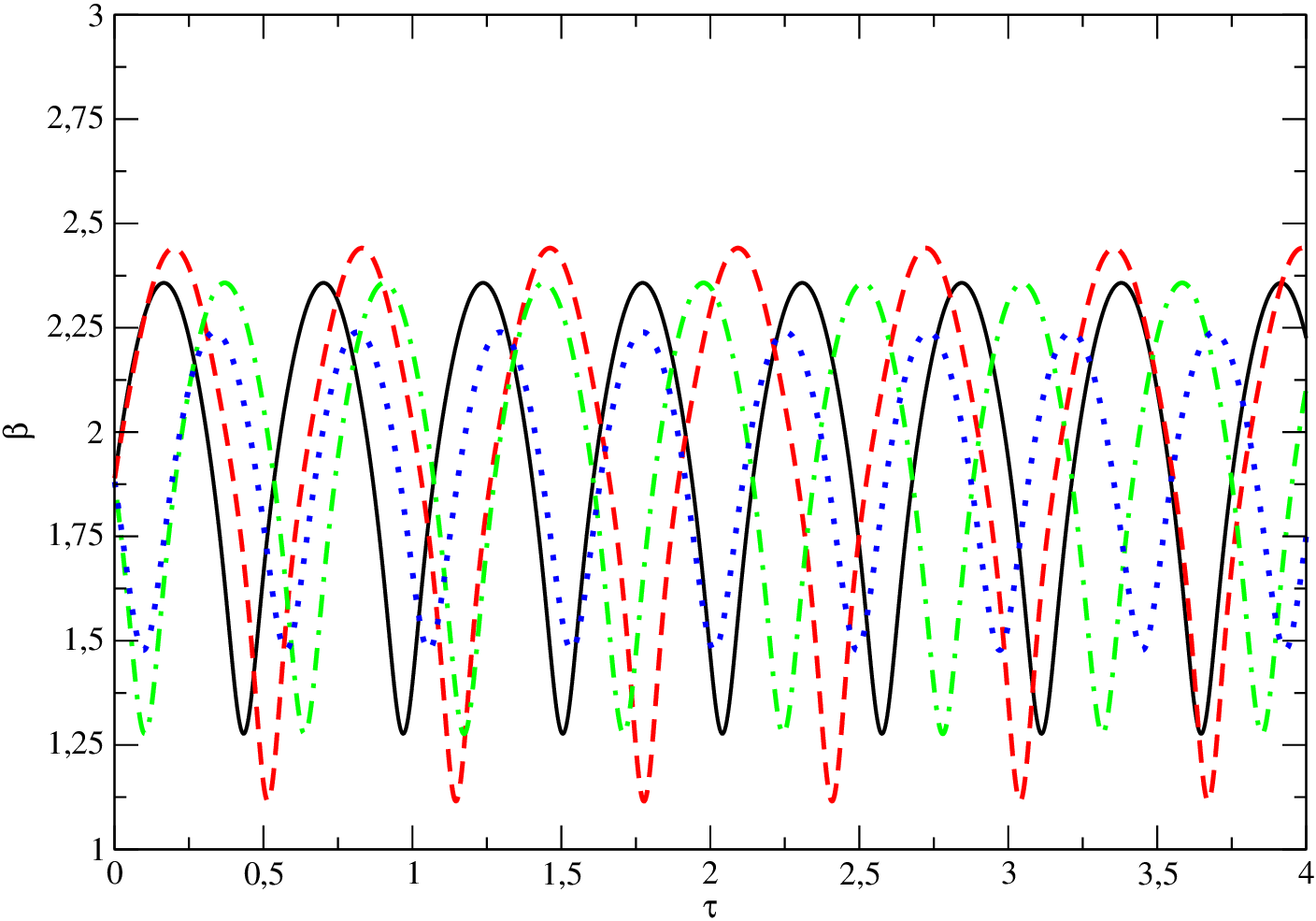}
\hspace{0.5cm}
\includegraphics[width=8.0cm,height= 9.0cm,angle=0]{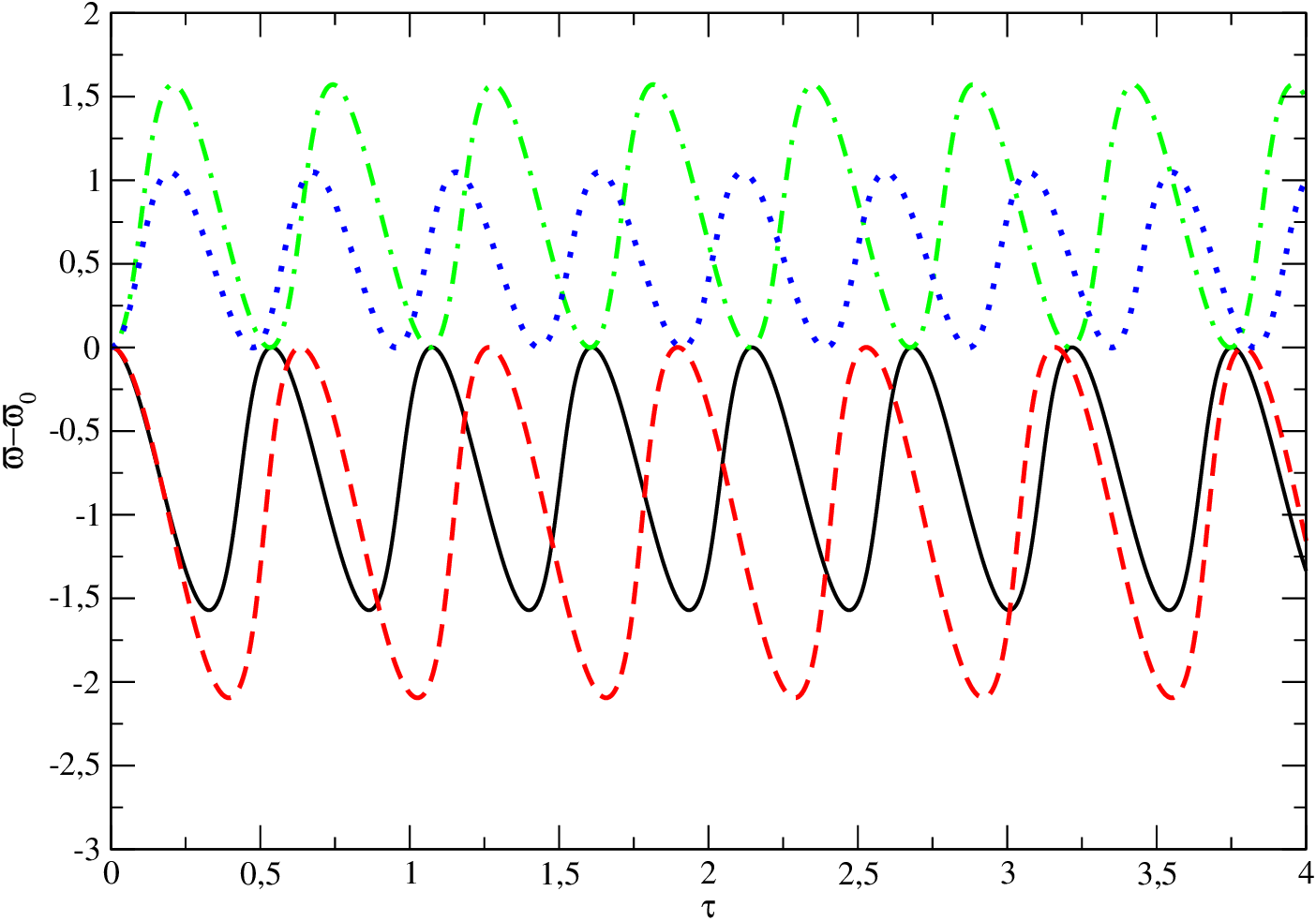}
\caption{{ The left panel shows the evolution of the inclination angle $\beta$ as a function of dimensionless  time $\tau$. The time $\tau=0$ corresponds to 
the system being on the critical curve. For these calculations, $q=1$, $e_0=0.5$ and $\beta_{0}={3{\rm \pi}/ 5}$, and $\sigma \approx 2$.
Curves of different line style correspond to different initial values of ${\hat \varpi},$ namely ${\hat \varpi_0}$, see the text for their description.
The right panel shows the  evolution of ${\hat \varpi} - {\hat \varpi_0}\equiv  \varpi -  \varpi_ 0$ for these calculations. Curves with the same line  style in each panel
correspond to the same calculation. }}
\label{nbeta}
\end{center}
\end{figure}

\begin{figure}
\begin{center}
\vspace{1cm}
\hspace{-1cm}
\includegraphics[width=8.0cm,height= 9.0cm,angle=0]{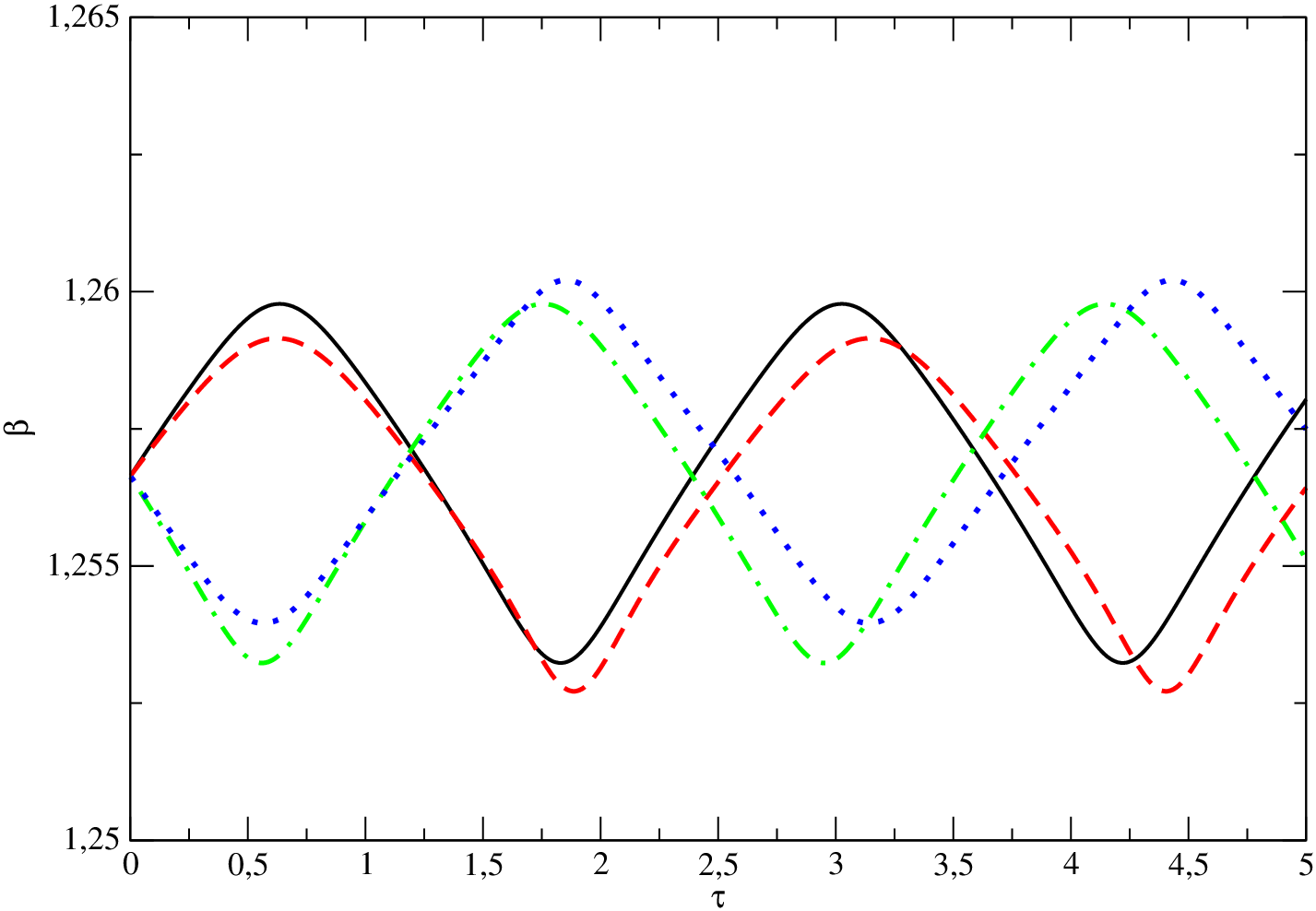}
\hspace{0.5cm}
\includegraphics[width=8.0cm,height= 9.0cm,angle=0]{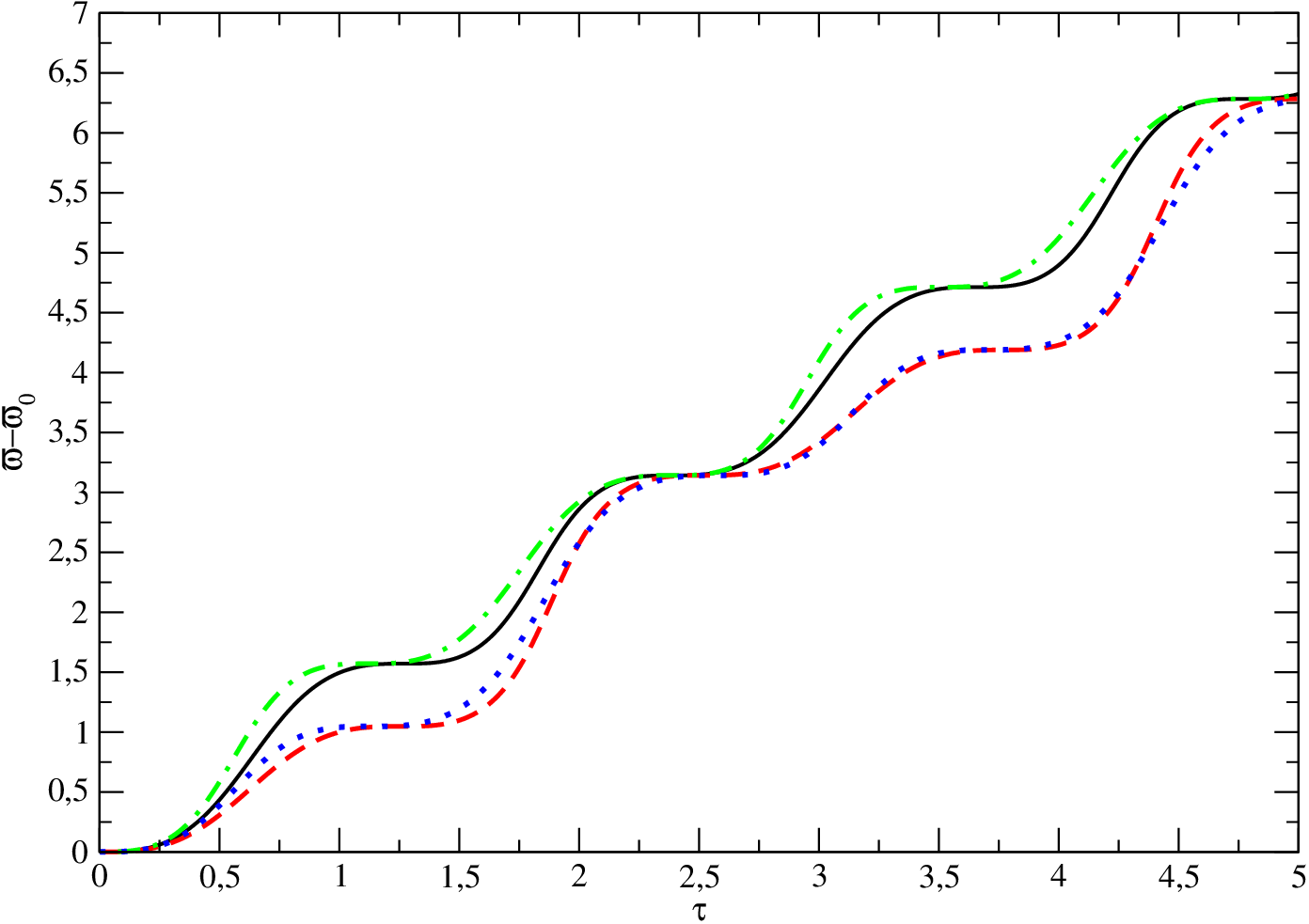}
\caption{{As for  Fig. \ref{nbeta}, but  in the left panel  we have the small mass ratio,
  $q=10^{-4},$ and prograde  stellar rotation with , $\beta={2{\rm \pi}/ 5}$. In this case, $\sigma\approx 5.31,$ for which there is a 
sharp maximum on the curve representing the estimated amplitude $\delta$ given by (\ref{libampl1})  ( see the left panel of Fig. \ref{nd_prog} {and the discussion in  Section \ref{ampest}}).
The right panel shows the corresponding evolution of ${\hat \varpi}-{\hat \varpi_0} \equiv  \varpi -  \varpi_ 0.$. {In this case ${\hat\varpi}$ does not librate}} }
\label{nbeta1}
\end{center}
\end{figure}

\onecolumn

{By appropriately differentiating the right hand side of equation (\ref{ev5}) after making use of equations (\ref{ev7})-(\ref{ev6})  and equation (\ref{ev81})
we readily obtain a somewhat lengthy expression for ${\cal D}$.  This takes the  form
\begin{align}
&{\cal D}= {(1+q)\over 3 q}{\sigma^2\cos\beta_0\sin\beta_0\over (1-e_0^2)^2}
+{\sigma {\tilde a_0}^{2}\sin \beta_0 \over 15\tilde I (1-e_0^2)^{3/2}}+\nonumber\\
&\frac{2{\tilde S}\sqrt{(1-e_0^2)}\sin\beta_0}{(1+{\tilde S}\cos\beta_0 (1-e_0^2)^{-1/2})}
\left( {\gamma_E(1+q)\tilde a_0^4\over q(1-e_0^2)^2}  +{ 52+50e_0^2+3e^4_0\over 8 (1-e_0^2)^6}\right.\nonumber\\
&\left. +{(1+q)\sigma^2(5\cos^2\beta_0 -1)\over 15q (1-e_0^2)^3}
 +{\sigma \tilde a_0^2\cos \beta_0 \over 10\tilde I (1-e_0^2)^{5/2}}\right)
\label{D}
\end{align}}

{From equations (\ref{ev2}) and (\ref{ev3}) it follows that the evolution equation for the angle $\beta$ can be represented
in the form
\begin{align}
&\hspace{0mm}{d \beta \over d\tau}=A_{\beta,e^2}\sin \beta \sin 2\hat \varpi, \hspace{2mm} {\rm where}\hspace{2mm}\nonumber\\
& A_{\beta,e^2}=
{3qe^2(1+{e^2/ 6})\sigma {\tilde a_{0}}^{-1}\over 5{\tilde I}
(1-e^2)^{9/2}}\left(1+\frac{(1+q)\cos\beta{\tilde I}{\tilde a_0}^{-2}\sigma}{q\sqrt{1-e^2}}\right).
\label{crit4} 
\end{align}
In the limit of small variation of $\beta$, in (\ref{crit4}) we set $d\beta/d\tau=d\delta/d\tau,$ $\sin\beta=\sin\beta_0,$ and  $A_{\beta, e^2} = A_{\beta_0,e_0^2}$. The latter
 is a constant as ${\tilde a}={\tilde a_0}$  and $\sigma$ are conserved.
We note that when the spin angular momentum is much less than the orbital angular momentum as is expected for $q$ of order unity $A_{\beta_0,e_0^2}$ is positive. However in the opposite case it can be negative. 
But, there is no restriction on $\hat\varpi$ and we see that the system is invariant
under the shift  $\hat\varpi\rightarrow  \hat\varpi+{\rm \pi}/2$ together with $A_{\beta_0,e_0^2}\rightarrow  -A_{\beta_0,e_0^2}.$
Thus without loss of generality we may set $A_{\beta_0,e_0^2}\rightarrow |A_{\beta_0,e_0^2}|.$
We remark that when $A_{\beta_0,e_0^2}=0,$  $\beta$ remains fixed at $\beta_0$ while ${\hat \varpi}$ is fixed at a value that can be chosen arbitrarily. We shall not consider this case further.}

{ For general $\beta_0$ it is seen that $\hat\varpi=\varpi_j = j{\rm \pi/2},  j=0, 1, 2, 3$ correspond to fixed points of the system which alternate between being stable and unstable.
 Note that a change from stability to instability and vice versa occurs when  ${\cal D}A_{\beta_0,e_0^2}\sin\beta_0\cos2\varpi_j$ changes sign. It is convenient to introduce a new time variable,\\  $\tau_{1}=A_{\beta_0,e_0^2}\tau$ and express (\ref{crit3}) and (\ref{crit4})
in the form
\begin{equation}
{d \hat \varpi \over d\tau_1}=-b\delta 
\label{crit41} 
\end{equation}
and 
\begin{equation}
{d \delta \over d\tau_1}=\sin \beta_0 \sin 2\hat \varpi,
\label{crit5} 
\end{equation}
where
\begin{equation}
b=\frac{{\cal D}}{A_{\beta_0,e_0^2}}
\label{crit6} 
\end{equation}}

\noindent {Equations (\ref{crit41}) and (\ref{crit5}) are  equivalent to a single second order differential  equation for $\hat\varpi$ as a function of  time}
\begin{equation}
{d^2 \hat \varpi \over d\tau_1^2}=-{b}\sin \beta_0 \sin (2\hat \varpi).
\label{crit9} 
\end{equation} 

It is convenient to rescale $\tau_1$ and $\delta$ to remove  {$b$} and $\sin \beta_0$ from (\ref{crit5}) and (\ref{crit9}) using
the substitution {$\tilde \delta =\delta\sqrt {| b|/\sin\beta_0}$, $\tilde \tau=\tau_1 \sqrt{|b|\sin\beta_0}$ }and 
bring (\ref{crit41}), (\ref{crit5})  and (\ref{crit9}) to the form
{\begin{equation}
{d \hat \varpi \over d\tilde \tau}=-{\rm sgn}(b) \tilde \delta, \quad 
{d \tilde \delta \over d\tilde \tau}=\sin  2\hat \varpi \quad 
{d^2 \hat \varpi \over d\tilde \tau^2}= -{\rm sgn}(b)\sin (2\hat \varpi).
\label{crit10}
\end{equation}  
As noted above we can change the sign of $\sin2{\hat \varpi}$ on the right hand side  of the last equation in (\ref{crit10}) by making the shift $\hat\varpi \rightarrow \hat\varpi +{\rm \pi}/2.$
Accordingly, we may take this sign to be negative without loss of generality.}
The last of eqns (\ref{crit10}) {is a standard pendulum equation. This} can be easily integrated to give
\begin{equation}
{\left({d\hat \varpi \over d\tilde \tau}\right)}^2- \cos (2\hat \varpi)=C,
\label{crit11} 
\end{equation} 
{For solutions that oscillate between $|\sin\hat \varpi_0|$ and $-|\sin \hat\varpi_0|,$
 $C=- \cos (2\hat \varpi_0)$. {Taking $\hat\varpi_0$ to be  the initial value of $\hat\varpi$} as stated above,
 \footnote{As the system is autonomous, for solutions with libration  we may choose a libration limit to  be ${\hat \varpi_0}$  and a time at which this occurs to be $t=0$
  without loss of generality.} 
then at  $t=0$, both $\tilde \delta$ and ${d\hat \varpi / d\tilde \tau}$ are 
equal to zero.}
Here we remark that the solutions with libration have the constant $C$ such that $|C|<1.$ Solutions with $C>1$ are such that ${\hat \varpi}$ circulates.
The amplitude of variation of $\delta$ is similar to that of  solutions librating with large amplitude when $C$ slightly exceeds unity, but it decreases
as $C$ increases ultimately leading to values expected from the discussion of Sections \ref{tidalest} and \ref{Einsteinest}.  

For solutions undergoing libration, the solution of (\ref{crit11}) is brought  into  standard form by the substitution $y={\sin \hat \varpi/
|\sin \hat \varpi_0|}$. 
Then, an implicit solution  can be expressed in terms of an incomplete elliptic integral 
of the first kind
\begin{equation}
\tilde \tau ={1\over \sqrt{2}}\int^{1}_{y}{dy^{'}\over {\sqrt{(1-{y^{'}}^2)(1-k^2{y^{'}}^2)}}},\hspace{2mm}{\rm where}\hspace{2mm}k=|\sin \hat\varpi_0|
\label{crit12} 
\end{equation} 
When $\sin\hat\varpi_0 > 0,$ equation (\ref{crit12}) describes the solution as $y$ oscillates between $1$ and $-1$. It subsequently retraces this moving between $-1$ and $1$, thereafter being periodic in $\tilde\tau$ with period $2\sqrt{2}K(k),$ where $K(k)$ is the complete elliptic integral of the first kind.
This evolution also applies when $\sin\hat\varpi_0 <0.$ Though in this case the solution starts with $y=-1$  and to describe the initial phase the sign
of the integral in (\ref{crit12}) is reversed.

From the analysis made above we can deduce a number of important consequences. Namely, the motion is
 periodic, with the period in time $t= t_*\tau_1$ being equal to
\begin{equation} 
P_{lib}={2\sqrt{2}t_* A^{-1}_{\beta_0,e_0^2} (\sin{\beta_0}|b|})^{-1/2}K(k),
\label{libperiod}
\end{equation}
and 
with a typical amplitude of variation of $\delta$ given by
\begin{equation}   
 |\delta/{\tilde \delta}| \sim  \sqrt{\sin \beta_0 / |b|}= \sqrt{A_{\beta_0,e_0^2}\sin\beta_0/{\cal D}}.
\label{libampl1}
\end{equation} 
The angle $\hat \varpi $ librates  around zero 
{\footnote{{As the system is invariant to shifting $\hat \varpi_0$ by a multiple of ${\rm \pi},$ the libration centre may also be shifted in this way.}}.} 
The amplitude of libration of ${\hat \varpi}$  {is $|{\hat \varpi_0}|$ with ${\tilde \delta}$ expected to be of order unity when this quantity is of order unity.
Accordingly, we set ${\tilde \delta}=1$ when using (\ref{libampl1}) to make estimates.}

\subsubsection{The amplitude of the variation in $\beta$ as a function of  
parameters of the problem}\label{ampest}
{ We represent the amplitude  of $\delta= \beta-\beta_0,$  given by equation (\ref{libampl1}) and evaluated on critical curves, in Figs. 
\ref{nd_prog}, and
 \ref{nd_retr2} 
 with input  parameters the same as those adopted
in Figs. \ref{na_prog}, 
and \ref{na_0_e05}, respectively.
 In particular, $e_0=0.5$ in all cases.
In what follows we simply denote this amplitude by ${\delta}.$
The left panel of Fig. \ref{nd_prog} illustrates the prograde case with $\beta_0= 2{\rm \pi}/5,$  where only curves corresponding to  small mass ratios, $q,$ in the range 
$10^{-5}-10^{-2}$ are shown. The right panel of Figs \ref{nd_prog} and the left panel of Fig. \ref{nd_retr2} illustrate the retrograde cases with $\beta_0=3{\rm \pi}/5,$  and $\beta_0=4{\rm \pi}/5$
respectively. The polar case with $\beta_0= {\rm \pi}/2$ is illustrated in the right panel of Fig.~\ref{nd_retr2}. 

One can see from Figs. \ref{nd_prog} and \ref{nd_retr2} that in general $\delta$ is smaller than unity and, therefore, the assumption of the smallness
of $\delta $ made for our analytical work is justified for most allowed parameters.
 
However, there are two
possible exceptions. Firstly, $\delta $ can be order of one  when rotation is retrograde and $q$ is sufficiently large,
$q\sim 1-10$, see the regions of the  solid and dashed curves in the right panel of Fig. \ref{nd_prog} and the left panel of Fig. \ref{nd_retr2} for $\sigma < \sim 1.$ 
This corresponds to  the lower branches of the corresponding critical curves as they approach  $\tilde a_{min}$ defined in eq. (\ref{lim1}) 
and, accordingly, the orbital periastron distance approaches the larger of the stellar radius or the tidal disruption radius . 

The second situation occurs when  when the mass ratio 
and for  some particular value of $\sigma,$,   $\delta$ has a sharp maximum, ( see e.g. the dot dashed and 
dot double dashed curves in the left panel of Fig. \ref{nd_prog}). 
This happens when the quantity ${\cal D}$ defined in eq. (\ref{D}) is zero for a prescribed value of $\sigma$. We illustrate this effect in Fig. \ref{ndel_prog}, where we plot the absolute value 
of $(\tilde a_{\cal D}(\sigma) -\tilde a_{0}(\sigma))/{\tilde a}_0(\sigma)$, where $\tilde a_{\cal D}(\sigma)$ is defined by the condition ${\cal D}(\tilde a_{\cal D},\sigma)=0$.
We note  that  here we do not display the dependence of ${\cal D},$ and ${\tilde a}_{\cal D}$ on quantities other than $\sigma$ as these are fixed.
The  curves plotted correspond to the prograde case illustrated in Figs \ref{na_prog} and \ref{nd_prog}. We see that the values of $\sigma$ for which the sharp maxima occur in the
former Figure correspond to the sharp minima in the latter. 
However, these sharp maxima are unrealistic because the variation of ${\cal D}$ with $\delta$ has not been taken into account.
Where ${\cal D}=0,$ the right hand side of (\ref{eccb}) should be replaced by $-(\partial{\cal D(\beta_{0})}/\partial {\beta_0}) \delta^2/2.$
Here we recall that $e_0^2$ is a function of $\beta_0$ through (\ref{initC}). Following this (\ref{crit41}) should be replaced by
\begin{equation}
{d \hat \varpi \over d\tau_1}=-b'\delta^2,\hspace{2mm} {\rm where}\hspace{2mm} b'= \frac{1}{2A_{\beta_0,e_0^2}}\frac{\partial {\cal D}}{\partial \beta_0}
\label{crit42} 
\end{equation}
From (\ref{crit5}) and (\ref{crit42}) it straight forward to obtain an estimate for the magnitude of $\delta$ given by
$\delta\sim |\sin\beta_0/b'|^{1/3}.$
Thus, extreme maxima do not occur and the wings on each side should connect smoothly as has been verified numerically
(see below). In fact the values of $|\delta|$ may dip if the magnitude of the derivative of ${\cal D}$ is large.

Note that there could also
be a situation where $\delta$ sharply tends to zero, see the blue dot dot dashed curve in the right panel of  Fig. \ref{nd_prog}. 
\noindent This happens when
the quantity in braces in the expression (\ref{crit4}) for $A_{\beta_0,e_0^2}$ is equal to zero for a  particular value
of $\sigma$.

\subsection{Numerical verification of analytic estimates}\label{numerical}

We present  numerical solutions of equations (\ref{ev2}) and (\ref{ev5}) in Figs. \ref{nbeta},
and \ref{nbeta1} 
 We use differential equation found by differentiating  eq. (\ref{ev9}) with respect to time  to provide another equation enabling 
 the determination of the  evolution of the  eccentricity. 
 We assume that initial values 
of the parameters  obtained by solving the  evolution equations 
 are such  that the system is initially 
on a critical curve. We consider two typical cases, where large variations of $\beta $ are expected from the discussion
in Section \ref{ampest}.

The first case has $q=1$ and retrograde rotation with $\beta_0={3{\rm \pi} / 5}.$
In addition  $e_0=0.5,$  $\sigma\approx 2$, and $\tilde a\approx 2.97$. 
It is illustrated it  in  Fig. \ref{nbeta}.
 The critical curve is  illustrated by  the {red} dashed curve in the right panel of  Fig. \ref{na_prog}.  The initial values of the run  belong to the lower branch 
 where  the rotational frequency is close to its maximum value.
 As seen from the corresponding
curve in  the right panel of Fig. \ref{nd_prog}, the analytic theory predicts  the amplitude of variations of, $\beta$, 
as given by equation (\ref{libampl1}) to be  $\sim 0.5$.
In Fig. \ref{nbeta}  solid, dashed, dot 
dashed and dotted curves are for different initial values of ${\hat \varpi}$,  namely ${\hat \varpi}_{0}={{\rm \pi} / 4}$, ${{\rm \pi} / 3}$, ${3{\rm \pi} / 4}$ and 
${5{\rm \pi}/ 6}$, respectively. 
As seen from Fig.  \ref{nbeta} 
 both angles, $\beta$ and ${\hat \varpi}$ exhibit periodic motion, with the characteristic amplitude of variations of $\beta$ being $\sim 0.5$ as expected. However, this quantity 
 depends on ${\hat \varpi_0}$. Also, variations of $\beta $ with respect to $\beta_0$  are asymmetric,  being larger for values of $\beta < \beta_0$.
On the other hand, the system spends somewhat longer periods of time with  $\beta > \beta_0$.
The difference ${\hat\varpi}-{\hat \varpi_0}$ is negative when ${\hat \varpi}_0 < {{\rm \pi} / 2}$ and positive otherwise. 
Interestingly, this case illustrates the  possibility of having  the evolution of the  system causing it to oscillate between prograde and retrograde states.

The second case we consider investigates the possibility of having a sharp resonance-like increase of the amplitude, $\delta,$ in the low mass
case near a point on the critical curve where ${\cal D}(\sigma)=0$.  
To illustrate this possibility we consider a calculation with   $q=10^{-4}$, $e_0=0.5$,
$\beta_{0}={2{\rm \pi} / 5},$ $\sigma\approx 5.31$ and $\tilde a\approx 11.04$. 
The results are presented in Fig. \ref{nbeta1} with line styles as in the previous Fig. 
As seen from the green dot dashed curve in the left panel of Fig.~\ref{nd_prog}, when the parameters are chosen in this way, 
even though the tidal interaction is relatively weak,
$\delta, $ is expected to be relatively large. 
The  numerical results shown in Fig. \ref{nbeta1} confirm this prediction, giving a 
typical amplitude of variations of $\beta$ order of one per cent.  
Though significant, in accordance with  expectations from 
the discussion in Section \ref{ampest},  the large and  extremely localised maximum seen in the left panel of  Fig. \ref{nd_prog} is absent.
Note that unlike the previous case  dependence of the  evolution of $\beta$ and ${\hat \varpi}$ on ${\hat \varpi}_{0}$ is practically absent. 
Also, the variation of $\beta$ relative to $\beta_0$  is symmetric.  


\section{Conclusions and Discussion}
\label{concl}     
{In this work we have developed and generalised results reported in IP  concerning the non-dissipative tidal evolution 
of the inclination angle between the  stellar  and  orbital angular momenta, $\beta.$ 
This is applicable to a binary system with a stellar primary and a compact
perturbing companion.
The evolution of $\beta$ is coupled  to the rate of  precession of the orbital  line of apsides
measured with respect to the line of nodes which itself precesses.
IP  considered only the classical contribution to apsidal motion due to tidal distortion \citep{St1939}. 
The extension to include the effects  due to  rotational distortion and  
Einstein precession, as well as the precession of the line of nodes mentioned above,
for arbitrary orbital eccentricity was discussed in Sections \ref{inclination} - \ref{periapsis}, with technical details supplied in appendix \ref{app}.
Section \ref{inclination} also included a brief review of the equation governing the evolution of $\beta$ derived in IP. 

This evolution is a qualitatively new effect, arising from the symmetry breaking of the tidal
bulge and associated gravitational field by rotation, leading to the appearance of a non dissipative torque acting between the primary star and orbit. 
Unlike the one leading to the usual precessional dynamics, this is directed in the plane containing the angular momentum vectors and so can change $\beta.$
Being non-dissipative,  both  the orbital and rotational energies are conserved, while conservation of angular momentum relates changes in $\beta$ to changes
in the eccentricity (see Section \ref{eqdim}).
In this paper  we  provided an extensive analytic treatment of this dynamics.

We remark that this evolution occurs even when one 
of the binary component is point-like, which makes it qualitatively  different  to that associated with the usual precessional 
dynamics driven by stellar flattening.  This could also lead to  changes in $\beta$, but only when both components are subject to the action of tidal forces, see e.g. \cite{PR} and references therein. 



In Section \ref{eveq} we considered  the situation when only one physical source of apsidal precession dominates over the others
 and provided  estimates for the expected typical change of the inclination angle $\beta$, $\Delta \beta.$
We also found  conditions on the parameters defining  the system  for one of these effects to dominate. 
We found  that, when all properties of the system are fixed apart from the ratio of the rotation frequency of the primary
to the orbital mean motion, $\sigma = \Omega_r/n_0$,
$\Delta \beta$ approaches its maximal  value when $\sigma$ increases sufficiently. 
This value is given by equation (\ref{ev17}). Estimates of $\Delta \beta$ which are valid for smaller values of $\sigma$  when  
different sources of apsidal motion dominate  are given by eqns. (\ref{ev15}) and (\ref{ev15a}).

In Sections \ref{S5} and \ref{evolution} we went on to consider the situation when  a solution of our dynamical system crosses a so-called 'critical curve',
on which the total apsidal precession rate is zero, for a given value of the inclination angle, $\beta_0$. 
In this case we expect that that variation of $\beta$ in the vicinity  will be much larger than was found in Section \ref{eveq} 
where  it  was assumed that one process dominated.
In Section \ref{S5} we provided an extensive analysis of the properties of  critical
curves finding that their  existence  is possible only when $\cos \beta_0 < {1/ \sqrt {5}}$. 

In Sections \ref{evolution} - \ref{numerical} we studied solutions of our dynamical system in the vicinity of these curves. 
We employed  an analytic approach, under the  simplifying assumption that $\Delta \beta $ is small, in Sections \ref{evolution} - \ref{ampest}. In addition, we performed direct numerical solutions of the dynamical equations describing our system for two representative cases, one with $q=1$ and one with $q=10^{-4},$ with the object of confirming our analytic estimates in Section \ref{numerical}.

Solutions in the vicinity of critical curves are periodic, they demonstrate several unusual features. 
Namely, unlike for the standard situation the apsidal angle can  change periodically (librate),
while for a strong enough interaction, $\Delta \beta$ could be large enough to change the rotation of the primary from being prograde to retrograde and vice versa.
   
In an accompanying paper \cite{IP1} we  provide, as an addition to the studies described here,
a preliminary numerical analysis of the parameter space of the problem for mass ratios $q=10^{-3}$, $q=1$ and $q=10^3$ and eccentricity $e_0=0.5$. We find that when $q=10^{-3}$ $\Delta \beta $ is always small and the regime of  critical curve crossing  is not found. However, when  $q=1$ or $q=10^3,$ the latter case expected to correspond to a primary of planetary mass orbiting a compact object of stellar mass, large variations  $\Delta \beta$ and the existence of a critical curve crossing  regime are found for a large range of the  parameters of the problem considered, provided that  $\sigma$ is large enough, and the  initial values $\beta_0 $ are larger than $\sim 1$.

Clearly, it is important to extend our results to the case of two tidally interacting components of a binary system as well as
take into account the possible role of other perturbing bodies. Also, as mentioned in  IP, the contribution of the toroidal component of the  displacement to the tidal response as well as  dynamic tides could be important. We intend to consider these issues within the framework of  formalisms developed by us in previous work,
  \citep[see e.g.][]{Papaloizou2005,IP7,Chernov2017}.}

\subsection{Potential applications}
{Finally,  the processes leading to apsidal motion and variability of the inclination between  orbital and spin 
angular momentum vectors discussed here
could  have applications to observations  of  eclipsing binaries  such as DI Herculis  \citep[e.g.][]{Sh,Alb} or transiting exoplanets on misaligned orbits \citep[see eg.][]{Alb1}.
Furthermore potentially misaligned hot and warm Jupiters  can be in orbits with significant eccentricity \citep{UM2022}.
{  A particular example is HD 80606 \citep[see eg.][]{Winn2009} which has $q\sim 4\times 10^{-3}$ and $e=0.93.$ 

In connection with such  exoplanet systems we note the following. As discussed above expected changed of
$\beta$ become  quite small when $q$ is small. Thus, while   apsidal motion could potentially  be  reversed or libration occur,
 changes to the orbital  inclination will be  small  when the stellar rotational axis is inclined with 
respect to the orbital plane. 

However, in the opposite limit $q\gg 1$ where it is assumed that tides operating in the exoplanet are more important than those 
acting on the central star,
 expected values of $\Delta \beta$ are  the same order  
as in case with  $q\sim 1$, see \cite{IP1} for  corresponding numerical examples.
 Thus, when a planet rotates sufficiently fast with its spin and orbital angular momentum misaligned, there may be a 
sizeable variation of $\beta$ with a small corresponding  variation of the inclination of orbital plane with respect to the line of sight on account of the  conservation of  angular momentum.
 Whether this possibility can be used to study rotational states of exoplanets for realistic parameters requires further study.}

{ \subsection{ The close binary DI Herculis}
 To provide an illustration  of  the expected variation of the inclination angle
 we consider  DI Herculis. The parameters of this system are given  in Table 1 of \cite{PR}. It consists of two stars with masses $M_{1,2}=2.68$ and $2.48M_{\odot}$, 
 radii $R_{1,2}=5.15$ and $4.25R_{\odot}$, with  rotation velocities at their  surfaces $v_{1,2}=\Omega_{r,1,2}R_{1,2}=122$ and $118 km/s.$
 The  orbital period $P_{orb}\approx 10d$ and eccentricity $e\approx 0.5$.
  The observed apsidal precession rate for this system $\dot \varpi_{DI} \approx 7\cdot
10^{-10}s^{-1}$ is a factor of  two  smaller than expected when  invoking only the contribution of Einstein precession.  This  is explained by the 
contribution of  rotationally induced terms \citep[see eg.][]{Sh}.     

DI Herculis is of course a system with two stars of comparable densities, where the stellar spin axes can evolve by the standard mechanism 
associated with the interaction of the two oblate,  separately  precessing,  stars as analysed in e.g. \cite{PR}. Moreover, our analysis above is not strictly speaking valid for such a system since we have assumed that one binary component is point-like. However, to make a crude estimate let us consider the secondary star as our primary star, since according to \cite{PR} it has inclination, $\beta_{2}\approx \pi/2,$
while noting that  the other star has $\beta_1$ within two standard deviations of $\pi/2.$
Adopting this assumption we find that our dimensionless semi-major axis and angular velocity can be estimated as $\tilde a\approx 13.5$ and $\sigma \approx 10$, respectively. Using these data we estimate $\dot \phi$ entering equation (\ref{ev14})
as $\dot \phi \approx 200$. Also, adopting $\tilde I=0.1$ we easily see that the total angular momentum is mainly determined by its orbital part and, accordingly, the first expression in (\ref{ev14}) should be used. This analysis leads to $\Delta \beta_1 \approx 3\cdot 10^{-2}$ and the corresponding evolution timescale $\pi/\dot \varpi_{DI}\approx 150yr$. This is much smaller than given by  the standard  mechanism, which gives $\Delta \beta_{1,2} \sim O(1)$ for the parameters they adopted, on a comparable timescale, see Fig. 6 of \cite{PR}\footnote{Note that the angles $\beta$ defined in \cite{PR} are approximately the same
as our $\beta$ only for systems viewed edge on, as e.g. DI Herculis.}.

However, this  analysis neglects the possibility of evolution near a critical curve and the libration of $\beta_1$ , which might be  expected for this system, since apsidal precessional frequencies of different physical origin and sign are expected to be comparable. Then  $\Delta\beta_1$  may be significantly larger than the above estimate. However,
an accurate treatment of this possibility requires extension of our formalism on the case of two bodies of comparable densities, which is beyond of the scope of the present paper.}                        

In addition the phenomena considered here  could  also potentially  modify orbital evolution on the longer  time scale associated with dissipative tidal evolution. This  is a problem  for  future work.}

\section*{Acknowledgments}
PBI was supported in part by the grant 075-15-2020-780 ’Theoretical and experimental studies
of the formation and evolution of extrasolar planetary systems and characteristics of exoplanets’
of the Ministry of Science and Higher Education of the Russian Federation.

\section{DATA AVAILABILITY}

There are no new data associated with this article.

\begin{appendix}

\section{Apsidal precession due to rotational flattening of the star}\label{app}

\cite{BOC}  derived a general expression for joint precession of orbital angular momentum and precession 
of the apsidal line in the form
\begin{equation}
\dot {\bf l}={\bf \Omega}\times {\bf l}, \quad \dot {\mbox{{\boldmath$\epsilon$}}}={\bf \Omega}\times {\mbox{{\boldmath$\epsilon$}}}, 
\label{a1}
\end{equation} 
where ${\bf l}$ is the unit vector in the direction of orbital angular momentum, ${\bf l}={\bf L}/L$, ${\mbox{{\boldmath$\epsilon$}}}$ is the unit
vector in the direction of periastron,
\begin{equation}
{\bf \Omega} ={\Omega}_1{\bf s}+{ \Omega}_2{\bf l}, \quad \Omega_1=\Omega_Q\cos \beta, \quad \Omega_2=\Omega_Q{1-5\cos^2\beta\over 2},
\label{a2}
\end{equation}
{ where we recall that,} 
 ${\bf s},$ is the unit vector in the direction of stellar spin, ${\bf s}={\bf S}/S$,  $\cos \beta =({\bf s}\cdot {\bf l}),$  $\sin \beta =|{\bf s}\times {\bf l}|$, and
\begin{equation}
\Omega_Q=-k_2(1+q){\sigma^2\over (1-e^2)^2}{\tilde a}^{-5}n_{0}.   
\label{a3} 
\end{equation}

{In order to obtain an explicit expression for the apsidal precession rate it is convenient to use the orthonormal 'orbital'  frame
specified by  (\ref{e5}) together 
with the law of conservation of angular momentum written in the form
\begin{equation}
J{\bf j}=L{\bf l}+S{\bf s}= {\rm constant}.    
\label{a5}
\end{equation}}

We project equations of motion (\ref{a1}) onto the {the basis vectors defining the
 frame specified by} (\ref{e5}), in particular, we define projections of ${\bf \Omega}$ onto
this frame, $\Omega_{i}=({\bf \Omega}\cdot {\bf e_{i}})$, where $i=x^{'}, y^{'}, z^{'}$.
{These are found to be  
\begin{align}
&\Omega_{x'} = (J/S)\Omega_1 \sin i = \Omega_1\sin\beta, \hspace{3mm} \Omega_{y'}=0,\hspace{3mm}
{\rm and }\hspace{3mm}\nonumber\\ 
&\Omega_{z'}~=~\Omega_2~+~\Omega_1\cos\beta.   \label{components}
\end{align}}

With the help of eq. (\ref{a1}) one can find the evolution law of the basis vectors 
and form the quantity $ {\bf e}_{x'}\cdot {\dot {\bf e}}_{y'} -{\bf e}_{y'}\cdot{\dot {\bf e}}_{x'}.$ This is  independent of $d i/dt$ and found to be
\begin {align}
{\bf e}_{x'}\cdot{\dot {\bf e}}_{y'} -{\bf  e}_{y'}\cdot{\dot {\bf e}}_{x'}=-2\Omega_{x'}\cot i \label{sub1}
\end{align}

The vector ${\mbox{{\boldmath$\epsilon$}}}$ is always perpendicular to ${\bf e}_{z^{'}}$. Accordingly, it can be represented in the form
\begin{equation}
{\mbox{{\boldmath$\epsilon$}}}=\cos \varpi {\bf e}_{x^{'}}+\sin \varpi {\bf e}_{y^{'}},
\label{a8}
\end{equation} 
and, from eq. (\ref{a1}) we have
\begin{equation}
\dot {{\mbox{{\boldmath$\epsilon$}}}}=\cos \varpi ({\Omega_{z^{'}}{\bf e}_{y^{'}}}-\Omega_{y^{'}}{\bf e}_{z^{'}})+\sin \varpi ({\Omega_{x^{'}}{\bf e}_{z^{'}}}-\Omega_{z^{'}}{\bf e}_{x^{'}}).
\label{a9}
\end{equation}

\noindent On the other hand, differentiating (\ref{a8}) we get
\begin{equation}
\dot {{\mbox{{\boldmath$\epsilon$}}}}=(\cos \varpi {\bf e}_{y^{'}}-\sin \varpi {\bf e}_{x^{'}})\dot \varpi 
+\cos \varpi \dot {\bf e}_{x^{'}}+\sin \varpi \dot {\bf e}_{y^{'}}.
\label{a10}
\end{equation} 
Now we equate (\ref{a9}) to (\ref{a10}) { and 
project the resulting equation on ${\bf e}_{x'}$ and ${\bf e}_{y'},$ 
while noting that these are orthogonal unit vectors.  Both projections give
\begin{align}
\dot{\varpi} =\Omega_{z'}+\frac{1}{2}({\bf e}_{x'}\cdot{\dot {\bf e}}_{y'} -{\bf e}_{y'}\cdot{\dot {\bf e}}_{x'})\label{AA}
\end{align}
respectively. 
Substituting (\ref{sub1}) into (\ref{AA}) we obtain }
\begin{equation}
\dot \varpi=\Omega_{z^{'}} -\Omega_{x^{'}}\cot i.  
\label{a11}
\end{equation} 
Equation (\ref{a11}) determines the apsidal precession rate.

Using eq. (\ref{components}) we see, that
\begin{equation}
\Omega_{z^{'}}=-\Omega_{Q}{(3\cos^2 \beta -1)\over 2}  
\label{a11a}
\end{equation} 
{and
\begin {equation}
\Omega_{x'} = (J/S)\Omega_Q \sin i\cos\beta.\label{Om1} 
\end{equation}}
\hspace{-1mm}Finally, we substitute (\ref{a11a}) and (\ref{Om1}),
in (\ref{a11}) to find an explicit expression 
for the apsidal\\  precession rate
\begin{equation}
\dot \varpi=-\Omega_{Q}\left({(3\cos^2 \beta -1)\over 2}+{J\over S}\cos \beta  {\cos i}\right).
\label{a15}
\end{equation}

\end{appendix}

\label{lastpage}

\end{document}